\documentclass[prd,nofootinbib,reprint,twocolumn,showpacs,superscriptaddress,floatfix]{revtex4-1}
\pdfoutput=1 
\usepackage{bm, color} 
\usepackage{amssymb,amsfonts,slashed,amsthm,amsmath, graphicx, soul, empheq, cancel}
\usepackage{hyperref}
\usepackage[caption=false]{subfig}

\begin{document}

\newcommand{\vev}[1]{ \left\langle {#1} \right\rangle }
\newcommand{\bra}[1]{ \langle {#1} | }
\newcommand{\ket}[1]{ | {#1} \rangle }
\newcommand{\eV}{ \ {\rm eV} }
\newcommand{\KeV}{ \ {\rm keV} }
\newcommand{\MeV}{\  {\rm MeV} }
\newcommand{\GeV}{\  {\rm GeV} }
\newcommand{\TeV}{\  {\rm TeV} }
\newcommand{\1}{\mbox{1}\hspace{-0.25em}\mbox{l}}
\newcommand{\Red}[1]{{\color{red} {#1}}}

\newcommand{\lmk}{\left(}  
\newcommand{\rmk}{\right)}
\newcommand{\lkk}{\left[}  
\newcommand{\rkk}{\right]}
\newcommand{\lhk}{\left \{ }  
\newcommand{\rhk}{\right \} }
\newcommand{\del}{\partial}  
\newcommand{\la}{\left\langle} 
\newcommand{\ra}{\right\rangle}
\newcommand{\half}{\frac{1}{2}}

\newcommand{\bea}{\begin{array}}
\newcommand{\eea}{\end{array}}
\newcommand{\beq}{\begin{eqnarray}}
\newcommand{\eeq}{\end{eqnarray}}
\newcommand{\eq}[1]{Eq.~(\ref{#1})}

\newcommand{\dd}{\mathrm{d}}
\newcommand{\Mpl}{M_{\rm Pl}}
\newcommand{\mg}{m_{3/2}}
\newcommand{\abs}[1]{\left\vert {#1} \right\vert}
\newcommand{\mphi}{m_{\phi}}
\newcommand{\Hz}{\ {\rm Hz}}
\newcommand{\for}{\quad \text{for }}
\newcommand{\Min}{\text{Min}}
\newcommand{\Max}{\text{Max}}
\newcommand{\Kahler}{K\"{a}hler }
\newcommand{\cphi}{\varphi}
\newcommand{\Tr}{\text{Tr}}
\newcommand{\diag}{{\rm diag}}

\newcommand{\SUf}{SU(3)_{\rm f}}
\newcommand{\Upq}{U(1)_{\rm PQ}}
\newcommand{\Zpq}{Z^{\rm PQ}_3}
\newcommand{\Cpq}{C_{\rm PQ}}
\newcommand{\ubar}{u^c}
\newcommand{\dbar}{d^c}
\newcommand{\ebar}{e^c}
\newcommand{\nubar}{\nu^c}
\newcommand{\Ndw}{N_{\rm DW}}
\newcommand{\Fpq}{F_{\rm PQ}}
\newcommand{\fpq}{v_{\rm PQ}}
\newcommand{\Br}{{\rm Br}}
\newcommand{\Lag}{\mathcal{L}}
\newcommand{\Lqcd}{\Lambda_{\rm QCD}}

\newcommand{\ji}{j_{\rm inf}} 
\newcommand{\jb}{j_{B-L}} 
\newcommand{\M}{M} 
\newcommand{\im}{{\rm Im} }
\newcommand{\re}{{\rm Re} }

\def\lrf#1#2{ \left(\frac{#1}{#2}\right)}
\def\lrfp#1#2#3{ \left(\frac{#1}{#2} \right)^{#3}}
\def\lrp#1#2{\left( #1 \right)^{#2}}
\def\REF#1{Ref.~\cite{#1}}
\def\SEC#1{Sec.~\ref{#1}}
\def\FIG#1{Fig.~\ref{#1}}
\def\EQ#1{Eq.~(\ref{#1})}
\def\EQS#1{Eqs.~(\ref{#1})}
\def\TEV#1{10^{#1}{\rm\,TeV}}
\def\GEV#1{10^{#1}{\rm\,GeV}}
\def\MEV#1{10^{#1}{\rm\,MeV}}
\def\KEV#1{10^{#1}{\rm\,keV}}
\def\blue#1{\textcolor{blue}{#1}}
\def\red#1{\textcolor{blue}{#1}}

\newcommand{\eff}{\Delta N_{\rm eff}}
\newcommand{\neff}{\Delta N_{\rm eff}}
\newcommand{\cc}{\Omega_\Lambda}
\newcommand{\Mpc}{\ {\rm Mpc}}
\newcommand{\Msolar}{M_\odot}


\title{High Quality QCD Axion via Electric-Magnetic Duality
}
\author{Shota Nakagawa, Yuichiro Nakai, Junxuan Xu, and Yufei Zhang
\\*[10pt]
{\it \normalsize Tsung-Dao Lee Institute, Shanghai Jiao Tong University, \\
No.~1 Lisuo Road, Pudong New Area, Shanghai, 201210, China} \\*[3pt]
{\it \normalsize School of Physics and Astronomy, Shanghai Jiao Tong University, \\
800 Dongchuan Road, Shanghai, 200240, China} 
}

\begin{abstract}
We propose a novel paradigm for the QCD axion with high-quality Peccei-Quinn (PQ) symmetry
on the basis of electric-magnetic duality in the conformal window of a supersymmetric gauge theory. 
PQ breaking fields, that contain the QCD axion, emerge in the magnetic theory
and possess a large anomalous dimension,
which leads to not only generation of an intermediate scale of spontaneous PQ breaking
but also significant suppression of explicit PQ symmetry breaking operators.
The high PQ quality and the absence of a Landau pole in the color gauge coupling
are achieved.
The parameter space to realize the correct abundance of the axion dark matter (DM) predicts
explicit PQ violation which may be probed by future measurements of the neutron electric dipole moment.
In the other viable parameter space, the lightest supersymmetric particle can become a DM candidate. 
Since the model naturally accommodates a mechanism to suppress the axion isocurvature fluctuation,
it provides a complete solution to the strong CP problem as well as the identity of DM.
\end{abstract}

\maketitle

\section{Introduction
\label{introduction}}

The strong CP problem is a major unsolved question in the Standard Model (SM) of particle physics.
Non-observation of the neutron electric dipole moment (EDM) puts a bound on the strong CP phase,
$|\bar{\theta}|\lesssim10^{-10}$ \cite{Baker:2006ts,Pendlebury:2015lrz}.
That is, CP violation is unnaturally suppressed in quantum chromodynamics (QCD).
The most popular solution to the problem is based on the Peccei-Quinn (PQ) mechanism \cite{Peccei:1977hh},
where a spontaneously broken global symmetry is introduced, resulting in the associated pseudo-Nambu-Goldstone mode,
called the axion \cite{Weinberg:1977ma,Wilczek:1977pj}.
Since the vacuum expectation value (VEV) of the axion field is set to a CP conserving vacuum below the QCD scale,
the strong CP phase $\bar{\theta}$ is cancelled out dynamically.
Interestingly, in the early Universe, the coherent oscillation of the same axion field behaves as a matter fluid
and can be a dominant component of dark matter (DM) \cite{Preskill:1982cy,Abbott:1982af,Dine:1982ah}.
The nature of the axion is controlled by the axion decay constant $f_a$ which is constrained as $10^8\GeV\lesssim f_a\lesssim 10^{12}\GeV$, where the lower bound comes from the astrophysical cooling argument \cite{Mayle:1987as,Raffelt:1987yt,Turner:1987by,Chang:2018rso,Carenza:2019pxu,Leinson:2014ioa,Hamaguchi:2018oqw,Leinson:2019cqv,Buschmann:2021juv}
and the upper bound is put by imposing a condition that the axion DM is not overproduced
for a natural initial amplitude of the axion oscillation.
Since the axion mass and coupling strength with SM particles depend on $f_a$,
various searches for the axion are currently ongoing and planned in a vast range of $f_a$
(for a summary of the current status, see Ref.~\cite{AxionLimits}).

While the axion is attractive as a solution to the strong CP problem and
a DM candidate, there still remain questions to be answered.
One is why spontaneous breaking of the PQ symmetry occurs at such an intermediate scale
hierarchically smaller than the Planck scale, inducing the instability of $f_a$ against radiative corrections
from higher scale physics.
Another issue is the quality of the global PQ symmetry
\cite{Dine:1986bg, Barr:1992qq, Kamionkowski:1992mf, Kamionkowski:1992ax, Holman:1992us, Kallosh:1995hi, Carpenter:2009zs, Carpenter:2009sw}.
In order to solve the strong CP problem,
any extra correction to the axion potential must be suppressed at the level of $\mathcal{O}(10^{-10})$,
which is incompatible with quantum gravity effects \cite{Kallosh:1995hi,Banks:2010zn,Witten:2017hdv,Harlow:2018jwu,Harlow:2018tng}.
A way to solve these two questions is to make the axion composite
\cite{Kim:1984pt,Choi:1985cb,Randall:1992ut,Izawa:2002qk,Yamada:2015waa,Redi:2016esr,DiLuzio:2017tjx,Lillard:2017cwx,Lillard:2018fdt,Gavela:2018paw,Lee:2018yak,Yamada:2021uze,Ishida:2021avk,Contino:2021ayn}.
The scale of $f_a$ is dynamically generated by dimensional transmutation
in a high-energy gauge theory with quarks that form the composite axion after the confinement,
and the theory does not allow dangerous interactions explicitly violating the PQ symmetry
so that the axion quality problem is ameliorated.
However, one general issue of a composite axion model is that since many new colored particles are introduced to the model, the SM color gauge coupling easily hits a Landau pole.
Another approach is the use of (super)conformal dynamics
\cite{Nakai:2021nyf,Nakagawa:2023shi}.
That is, a PQ breaking scalar field couples to a conformal field theory (CFT) and
holds a large anomalous dimension,
which leads to a significant suppression of PQ-violating higher-dimensional operators expected from quantum gravity effects.
A marginally-relevant operator triggers spontaneous breaking of the PQ symmetry at an intermediate scale.
The conformal invariance is also spontaneously broken, generating a mass gap in the CFT.
Although this model can significantly ameliorate the axion quality problem,
its complete solution is only achieved in a limited parameter space with a small gravitino mass.\footnote{
There have been several other attempts to the axion quality problem such as
warped extra dimension models~\cite{Flacke:2006ad,Cox:2019rro,Bonnefoy:2020llz,Lee:2021slp},
the visible axion
\cite{Rubakov:1997vp,Berezhiani:2000gh,Hook:2014cda,Fukuda:2015ana,Gherghetta:2016fhp,Dimopoulos:2016lvn,Gherghetta:2020ofz,Alves:2017avw,Liu:2021wap,Girmohanta:2024nyf}
and the introduction of a gauge symmetry protecting the PQ symmetry \cite{Cheng:2001ys,Harigaya:2013vja,Fukuda:2017ylt,Fukuda:2018oco,Ibe:2018hir,Choi:2020vgb,Yin:2020dfn,Chen:2021haa}.
}

In the present paper, we explore a new paradigm that a high-quality axion emerges through electric-magnetic duality
in an $\mathcal{N}=1$ supersymmetric non-Abelian gauge theory
\cite{Seiberg:1994pq}:
two different (electric and magnetic) gauge theories describe the same long distance physics.\footnote{
{The electric-magnetic duality in the axion electrodynamics has been recently discussed in Ref.~\cite{Csaki:2024plt}
in the context of the $\mathcal{N}=2$ Seiberg-Witten theory
\cite{Seiberg:1994rs}.} }
What is amazing in this duality is that the weakly interacting region of one theory is mapped
to the strongly interacting region of the other.
To build a high-quality axion model, we define our electric theory in conformal window at a UV scale.
Then, the theory flows into a conformal fixed point
where the magnetic theory gives a better description
and contains PQ breaking fields.
As we will see, the generation of the scale $f_a$ and the axion quality problem are addressed by
a combination of mechanisms realized in the composite and conformal axion models. 
Hence, the model occupies an intermediate position between the composite and conformal axions:
a larger number of color in the electric theory, which tends to suffer from a Landau pole in the ordinary QCD gauge coupling,
would lead to the composite axion, while a smaller number of color would make the electric theory a better description,
realizing the conformal axion
(in this case, we need to introduce elementary PQ breaking fields).
Then, the problems in the composite and conformal axion models are potentially resolved.
In fact, we will see that the high PQ quality and the absence of a Landau pole in the color gauge coupling are achieved at the same time.
Furthermore, it will be shown that the model possesses an intrinsic mechanism to suppress
the isocurvature fluctuation of the light axion and significantly relax a constraint on the scale of inflation.

The rest of the paper is organized as follows.
In \SEC{sec:Model}, we present our axion model in both electric and magnetic pictures,
and discuss the generation of the PQ symmetry breaking scale.
The stabilization of the saxion direction with supersymmetry (SUSY) breaking is also performed.
\SEC{sec:Landau} studies the issue of a Landau pole in the QCD gauge coupling.
In \SEC{sec:quality}, we estimate the quality of the PQ symmetry
and identify a viable parameter space of the model.
\SEC{sec:DM} discusses implications on the DM in our model
and the issue of isocurvature fluctuations.
Finally, \SEC{sec:Discussion} is devoted to conclusions and discussions.

\begin{table*}[t]
\vspace{0mm}
\centering
\begin{tabular}{c|c|c|c|c|c|c|c|c}
& $Q_a$ & $\bar{Q}^a$ & $Q_\alpha$ & $\bar{Q}^\alpha$ & $\eta_a^{~b}$ & $\xi_\alpha^{~\beta}$ & $Z_a^{~\alpha}$ & $\bar{Z}_\alpha^{~a}$ \\
\hline
$SU(N)$ & $\square$ & $\bar{\square}$ & $\square$ & $\bar{\square}$ & ${\bm 1}$ & ${\bm 1}$ & ${\bm 1}$ & ${\bm 1}$ \\
$SU(N_F/2)_1$ & $\square$ & $\bar{\square}$ & ${\bm 1}$ & ${\bm 1}$ & {\bf Adj} & ${\bm 1}$ & $\square$ & $\bar{\square}$ \\
$SU(N_F/2)_2$ & ${\bm 1}$ & ${\bm 1}$ & $\square$ & $\bar{\square}$ & $\bm 1$ & {\bf Adj} & $\bar{\square}$ & $\square$ \\
$U(1)_{\rm PQ}(\supset Z_{\widetilde{N}})$ & $\chi_1$ & $\chi_2$ & $-\chi_2$ & $-\chi_1$ & $-1$ & $1$ & $0$ & $0$ \\
$U(1)_{R}$ & $\frac{N_F-N}{N_F}$ & $\frac{N_F-N}{N_F}$ & $\frac{N_F-N}{N_F}$ & $\frac{N_F-N}{N_F}$ & $\frac{2N}{N_F}$ & $\frac{2N}{N_F}$ & $\frac{2N}{N_F}$ & $\frac{2N}{N_F}$ \\
\end{tabular}
\vspace{1mm}
\caption{The charge assignments in the electric theory.}
\label{tab:charge}
\end{table*}

\section{The Model
\label{sec:Model}}

We define our axion model with its electric picture and then discuss the magnetic dual. 
The generation of the PQ symmetry breaking scale is described by using the magnetic theory.
The axion couplings to gluons and photons are also presented.
Finally, we introduce SUSY breaking and perform the saxion stabilization.

\subsection{Electric theory}

Let us consider a supersymmetric $SU(N)$ gauge theory with $N_F$ pairs of vectorlike quarks, $Q$ and $\bar{Q}$,
which belong to the (anti-)fundamental representation of $SU(N)$. 
The theory is in conformal window and flows into a nontrivial IR fixed point
\cite{Intriligator:2007cp}. 
Here, our focus is on the case that the number of flavors is an even number,
parameterized as $N_F=2N-\delta$ with $\delta=2,4,6, \dots (<N/2)$. 
We then divide the quarks $Q,\bar{Q}$ into two parts, represented by $Q_a (\bar{Q}^a), Q_\alpha (\bar{Q}^\alpha)$ with $a \in (1,\dots,N_F/2)$ and $\alpha$ $\in (N_F/2+1, \dots,N_F)$,
and assign them with different charges under the global $U(1)_{\rm PQ}$ symmetry.
In addition, we introduce gauge-singlet chiral supermultiplets, $\eta_a^{~b}, \xi_\alpha^{~\beta}, Z_a^{~\alpha},\bar{Z}_\alpha^{~a},$
to stabilize extra fields appearing in the magnetic picture, which we will see later.
The theory respects $U(1)_{\rm PQ}\times SU(N_F/2)_1\times SU(N_F/2)_2$ flavor symmetry.
The charge assignments are summarized in Table \ref{tab:charge}
where we define PQ charges,
\beq
\chi_1 \equiv \frac{2N-N_F}{2N} \, ,
\qquad \chi_2 \equiv \frac{N_F}{2N} \, ,
\eeq
so that the $U(1)_{\rm PQ}$ is not anomalous under $SU(N)$ to avoid the axion coupling to
the $SU(N)$ gauge field.
The assignment of the $U(1)_R$ charge is determined by the superconformal nature.
The flavor symmetry $SU(N_F/2)_1$ is weakly gauged and the SM color gauge group $SU(3)_c$ as well as
the hypercharge $U(1)_Y$ is embedded in it.
The $U(1)_{\rm PQ}$ symmetry has anomaly under the $SU(3)_c$ whose anomaly coefficient
is given by $A_{U(1)_{\rm PQ} - SU(3)_{c} - SU(3)_{c}} = - \widetilde{N}$ with $\widetilde{N}\equiv N_F-N$.
Note that an anomaly-free (or gaugeable) discrete symmetry $Z_{\widetilde{N}}$ remains as a residual symmetry.
We impose the discrete symmetry to forbid some explicit $U(1)_{\rm PQ}$ violating operators
and realize the PQ symmetry at the renormalizable level.

The symmetry allows the following superpotential with a Planck-suppressed operator:
\beq
W_{\rm el} &=& \nu_1\eta_{a}^{~b}\bar{Q}^aQ_b + \nu_2\xi_{\alpha}^{~\beta}\bar{Q}^{\alpha}Q_\beta \nonumber\\ 
&+& Z_b^{~\alpha}\bar{Q}^bQ_{\alpha} + \bar{Z}_{\alpha}^{~b}\bar{Q}^{\alpha}Q_b + MZ_a^{~\alpha}\bar{Z}_{\alpha}^{~a}\nonumber\\
&-& \frac{\iota}{\Mpl} (\bar{Q}^a Q_a) (\bar{Q}^\alpha Q_\alpha) \, ,
\label{Wel}
\eeq
where $\nu_{1}, \nu_2, \iota$ are dimensionless coupling constants, $M$ is a mass parameter
and $\Mpl \equiv 1/\sqrt{8\pi G}$ denotes the reduced Planck mass scale with $G$ the Newton constant.
We have omitted to explicitly write coupling constants in front of the first and second terms in the second line.
As we will see, the Planck-suppressed operator plays an essential role for spontaneous breaking of the $U(1)_{\rm PQ}$ symmetry.
At the scale of $M$, the fields $Z,\bar{Z}$ are integrated out, and we obtain the low-energy effective superpotential,
\beq
W_{\rm el}^{(\rm eff)} &=& \nu_1\eta_{a}^{~b}\bar{Q}^aQ_b + \nu_2\xi_{\alpha}^{~\beta}\bar{Q}^{\alpha}Q_\beta  \nonumber\\ 
&-& \frac{\kappa}{M} (\bar{Q}^\alpha Q_a) (\bar{Q}^a Q_\alpha)
-\frac{\iota}{\Mpl} (\bar{Q}^a Q_a) (\bar{Q}^\alpha Q_\alpha) \, ,\nonumber\\
\label{Weleff}
\eeq
where $\kappa$ is introduced as a dimensionless coupling constant.

\begin{table*}[t]
\vspace{0mm}
\centering
\begin{tabular}{c|c|c|c|c|c|c|c|c|c|c}
& $q^a$ & $\bar{q}_a$ & $q^\alpha$ & $\bar{q}_\alpha$ & $\Phi+\Psi$ & $\bar{\Phi} + \bar{\Psi}$ & $\Sigma$ & $\Xi$ & $\eta_a^{~b}$ & $\xi_\alpha^{~\beta}$ \\
\hline
$SU(N_F-N)$ & $\square$ & $\bar{\square}$ & $\square$ & $\bar{\square}$ & ${\bm 1}$ & ${\bm 1}$&${\bm 1}$ &${\bm 1}$ & ${\bm 1}$ & ${\bm 1}$  \\
$SU(N_F/2)_1$ & $\bar{\square}$ & $\square$ & ${\bm 1}$ & ${\bm 1}$ & ${\bm 1}$+{\bf Adj} & ${\bm 1}$ & $\square$ & $\bar{\square}$ & {\bf Adj} & ${\bm 1}$ \\
$SU(N_F/2)_2$ & ${\bm 1}$ & ${\bm 1}$ & $\bar{\square}$ & $\square$ & ${\bm 1}$ & ${\bm 1}$+{\bf Adj} & $\bar{\square}$ & $\square$ & $\bm 1$ & {\bf Adj} \\
$U(1)_{\rm PQ}$ & $-1$ & $0$ & $0$ & $1$ & $1$ & $-1$ & $0$ & 0 & $-1$ & $1$ \\
$U(1)_{R}$ & $\frac{N}{N_F}$ & $\frac{N}{N_F}$ & $\frac{N}{N_F}$ & $\frac{N}{N_F}$ & $\frac{2(N_F-N)}{N_F}$ & $\frac{2(N_F-N)}{N_F}$ & $\frac{2(N_F-N)}{N_F}$ & $\frac{2(N_F-N)}{N_F}$ & $\frac{2N}{N_F}$ & $\frac{2N}{N_F}$ \\
\end{tabular}
\vspace{1mm}
\caption{The charge assignments in the magnetic theory.}
\label{tab:mSQCD charge}
\end{table*}

\subsection{Magnetic theory}

Corresponding to the presented electric $SU(N)$ gauge theory with $N_F\geq N+2$ flavors,
there exists the dual magnetic $SU(\widetilde{N})$ gauge theory \cite{Seiberg:1994pq},
which contains $N_F$ vectorlike pairs of dual quarks $q,\bar{q}$ and $N_F^2$ gauge singlets $\mathcal{M}_i^{~j}$ with $i,j=1,2,\cdots N_F$, as well as $\eta_a^{~b}, \xi_\alpha^{~\beta}$ introduced in the electric theory.
The superpotential includes $W_{\rm mag} \supset q \mathcal{M} \bar{q}$.
Note that the magnetic dual theory is also in conformal window.
According to the flavor structure of our model,
the gauge singlets $\mathcal{M}$ can be decomposed in terms of $N_F/2\times N_F/2$ matrices:
\beq
\mathcal{M} \equiv \left(
  \begin{array}{cc}
    \Phi\delta_a^b + \Psi_a^{~b} & \Sigma_a^{~\beta} \\
    \Xi_\alpha^{~b}  & \bar{\Phi}\delta_\alpha^\beta + \bar{\Psi}_\alpha^{~\beta} \\
  \end{array}
\right).
\eeq
Here, $\Phi \, (\bar{\Phi})$ and $\Psi \, (\bar{\Psi})$ are defined as the trace and traceless parts of
$\mathcal{M}_a^{~b} \, (\mathcal{M}_\alpha^{~\beta})$, respectively.
Table \ref{tab:mSQCD charge} summarizes the charge assignments in the magnetic dual theory.
We note that the `t Hooft anomaly
\cite{tHooft:1979rat}
is matched between the electric and magnetic theories.
In particular, the $U(1)_{\rm PQ}$ symmetry has anomaly under $SU(3)_c$
with the same anomaly coefficient as  the electric theory, $A_{U(1)_{\rm PQ} - SU(3)_{c} - SU(3)_{c}} = - \widetilde{N}$.

The correspondence between the gauge singlets $\mathcal{M}$ in the magnetic theory
and the quark bilinears in the electric theory is written as\footnote{
The electric baryon $B_{a_1 \dots a_k \alpha_1\dots \alpha_{N-k}}$
corresponds to the magnetic baryon $b_{a_{k+1}\dots a_{N_F/2} \alpha_{N-k+1} \dots \alpha_{N_F/2}}$.} 
\beq
(\Phi+\Psi)_{a}^{~~b} \leftrightarrow \frac{\bar{Q}^{b}Q_a}{\Lambda'} \, ,&&
(\bar{\Phi}+\bar{\Psi})_{\alpha}^{~~\beta} \leftrightarrow \frac{\bar{Q}^{\beta}Q_\alpha}{\Lambda'} \, ,\nonumber\\[1ex]
\Sigma_{a}^{~~\beta} \leftrightarrow \frac{\bar{Q}^{\beta}Q_a}{\Lambda'} \, ,&&
\Xi_{\alpha}^{~~b} \leftrightarrow \frac{\bar{Q}^{b}Q_\alpha}{\Lambda'} \, .
\label{MQQ}
\eeq
Here, the normalization factor $\Lambda'$ is a holomorphic and charge neutral scale and
given in terms of the holomorphic scale in the electric theory $\Lambda_{\rm el}$
and that in the magnetic theory $\Lambda_{\rm mag}$:
\beq
\Lambda'^{N_F}=(-1)^{N-N_F}\Lambda_{\rm el}^{3N-N_F}\Lambda_{\rm mag}^{3(N_F-N)-N_F} \, . 
\label{scale_mu}
\eeq
In the following discussion, we assume $\Lambda'\simeq\Lambda_{\rm el}\simeq\Lambda_{\rm mag}$. 
Then, corresponding to the superpotential of the electric theory in \EQ{Weleff},
the superpotential of the magnetic theory is given by
\beq
W_{\rm mag} &=& \lambda (\Phi q^a \bar{q}_a + q^a \Psi_a^{~b} \bar{q}_b) + \bar{\lambda} (\bar{\Phi} q^\alpha \bar{q}_\alpha 
+ q^\alpha \bar{\Psi}_\alpha^{~\beta} \bar{q}_\beta)\nonumber\\
&+& \rho q^a \Sigma_a^{~\beta} \bar{q}_\beta 
+ \bar{\rho} q^\alpha \Xi_\alpha^{~b} \bar{q}_b
- M_\Phi \Phi\bar{\Phi} -M_{\Sigma\Xi}\Sigma_a^{~\alpha}\Xi_{\alpha}^{~a}\nonumber\\
&+& \nu_1\Lambda'\eta_a^{~b}\Psi_b^{~a} + \nu_2\Lambda'\xi_\alpha^{~\beta}\bar{\Psi}_{\beta}^{~\alpha} \, , \label{Wmag}
\eeq
where $\lambda,\bar{\lambda},\rho,\bar{\rho}$ denote constant coefficients,
and we have introduced new dimensionful parameters, 
\beq
M_\Phi\equiv \iota\lmk\frac{N_F}{2}\rmk^2\frac{\Lambda'^2}{\Mpl} \, , \qquad 
\label{MPhi}
M_{\Sigma\Xi} \equiv \kappa\frac{\Lambda'^2}{M} \, .
\eeq
We can see from \EQ{Wmag} that the fields $\Sigma, \Xi$ form a mass term,
and $\Psi,\bar{\Psi}$ form mass terms with $\eta,\xi$, respectively.
After integrating out those massive fields, the effective superpotential is given by
\beq
W_{\rm mag}^{\rm (eff)}
&=& \lambda \Phi q^a \bar{q}_a + \bar{\lambda} \bar{\Phi} q^\alpha \bar{q}_\alpha 
- M_\Phi \Phi\bar{\Phi} \, .
\label{integrate}
\eeq
When $\Phi, \bar{\Phi}$ develop nonzero VEVs as we will see below, the $U(1)_{\rm PQ}$ symmetry is spontaneously broken,
and the $\Phi$ interaction with $q^a,\bar{q}_a$ induces the Chern-Simons term with the $SU(3)_c$ gluon
in the low-energy effective theory.
Thus, our model works as a KSVZ-like QCD axion model \cite{Kim:1979if,Shifman:1979if}, with the domain wall number $\widetilde{N}\equiv N_F-N$.

\subsection{Peccei-Quinn scale}

Based on the magnetic description of our model,
let us now discuss spontaneous breaking of the $U(1)_{\rm PQ}$ symmetry.
In addition to \EQ{integrate},
a non-perturbative effect of supersymmetric QCD (SQCD) generates a superpotential for $\Phi, \bar{\Phi}$
\cite{Nakagawa:2023shi}. 
Assuming $\Phi,\bar{\Phi}$ obtain nonzero VEVs, one can see that the dual quarks
$q^a, \bar{q}_a$ and $q^\alpha, \bar{q}_\alpha$ acquire masses of $\lambda \Phi$ and $\bar{\lambda} \bar{\Phi}$, respectively.
Then, at the energy scale of those quark masses, they are decoupled.
The low-energy effective theory is given by a pure super Yang-Mills
which leads to gaugino condensation,
\beq
W_{\rm gaugino} = (N_F-N) \Lambda_{\rm new}^3 \, , \label{gaugino}
\eeq
with
\beq
\Lambda_{\rm new}^3 = (\lambda\Phi)^{\frac{N_F}{2(N_F-N)}} (\bar{\lambda}\bar{\Phi})^{\frac{N_F}{2(N_F-N)}} \Lambda_{\rm mag}^{\frac{2N_F-3N}{N_F-N}},
\label{gaugino2}
\eeq
determined by matching the dynamical scales before and after integrating out $q, \bar{q}$.
The dynamically generated superpotential \eqref{gaugino} and the third mass term in \EQ{integrate} lead to the $\mathcal{F}$-term potential for the scalar components of $\Phi$ and $\bar{\Phi}$,
\beq
&&V_{\mathcal{F}} (\Phi, \bar{\Phi})\cdot Z_{\Phi} \nonumber\\
&=& 
\left| M_\Phi \Phi - \frac{N_F}{2} \bar{\lambda} \Lambda_{\rm mag}^{\frac{2N_F-3N}{N_F-N}} \lmk \lambda \Phi \rmk^{\frac{N_F}{2(N_F-N)}} \lmk \bar{\lambda} \bar{\Phi} \rmk^{\frac{2N-N_F}{2(N_F-N)}} \right|^2 
\nonumber\\
&+& 
\left| M_\Phi \bar{\Phi} - \frac{N_F}{2} \lambda \Lambda_{\rm mag}^{\frac{2N_F-3N}{N_F-N}} \lmk \lambda \Phi \rmk^{\frac{2N-N_F}{2(N_F-N)}} \lmk \bar{\lambda} \bar{\Phi} \rmk^{\frac{N_F}{2(N_F-N)}} \right|^2.\nonumber\\
&&
\label{VF}
\eeq
{Here, $Z_\Phi$ denotes the wavefunction renormalization factor for $\Phi, \bar{\Phi}$.}
The $\mathcal{F}$-term potential has two degenerate minima: 
\beq
 \la \Phi \ra &=& \la \bar{\Phi} \ra =0 \, ,
 \label{PQsymm}
\eeq
and
\beq
 \la \Phi \bar{\Phi} \ra &=& \frac{\Lambda_{\rm mag}^2}{\lambda \bar{\lambda}} \lmk \frac{2M_\Phi}{N_F \lambda \bar{\lambda} \Lambda_{\rm mag}} \rmk^{\frac{2(N_F-N)}{2N-N_F}}.
\label{flatdir}
\eeq
While the former vacuum indicates the symmetric phase,
the latter vacuum spontaneously breaks the $U(1)_{\rm PQ}$ symmetry.
We assume the latter vacuum is realized throughout the history of the Universe.

\begin{figure}[!t]
\centering
\includegraphics[width=8.5cm]{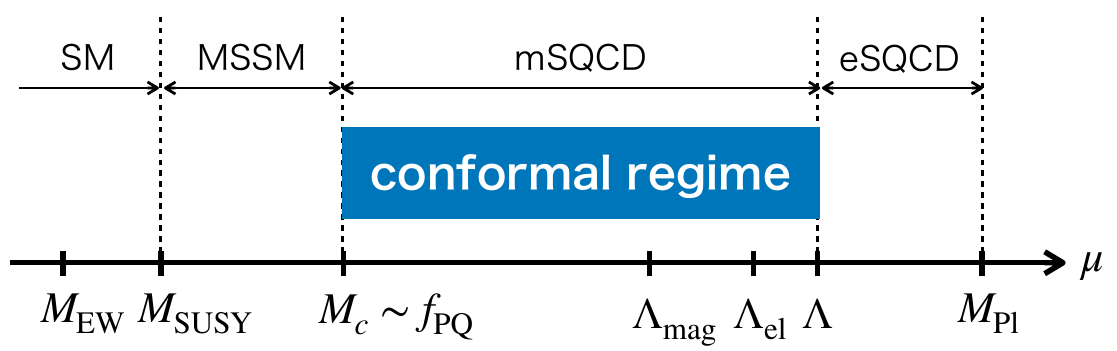}
\caption{
Hierarchies of scales in the model. 
Here, $M_{\rm SUSY}$ and $M_{\rm EW}$ represent the SUSY and electroweak symmetry breaking scales, respectively.
eSQCD and mSQCD denote the electric and magnetic pictures of our model.}
\label{fig:hierarchy}
\end{figure}

During the regime where the theory is at the conformal fixed point,
the kinetic terms of $\Phi, \bar{\Phi}$ experience a significant wavefunction renormalization
due to a large anomalous dimension,
which one needs to take into account to find the $U(1)_{\rm PQ}$ breaking scale
\cite{Nakai:2021nyf}.
The wavefunction renormalization factors for $\Phi, \bar{\Phi}$ and $q, \bar{q}$ are respectively given by
\beq
Z_\Phi &=& \left(\frac{M_c}{\Lambda}\right)^{-\gamma_\Phi},\\[1ex]
Z_q &=& \left(\frac{M_c}{\Lambda}\right)^{-\gamma_q},
\label{wfrenorm}
\eeq
with $\gamma_\Phi\equiv 4-6N/N_F$ and $\gamma_q \equiv -(2N_F-3N)/N_F$.
Here, $\Lambda$ stands for the scale at which the theory enters into the conformal regime
while $M_c$ represents the energy scale at which the theory exits from that regime. 
The canonically normalized $U(1)_{\rm PQ}$ breaking fields $\hat{\Phi}, \hat{\bar{\Phi}}$
and dual quarks $\hat{q}, \hat{\bar{q}}$ are written as
\beq
 \hat{\Phi} = 
 \sqrt{Z_\Phi} \, \Phi \, ,
 &&~~
 \hat{\bar{\Phi}} =  
 \sqrt{Z_\Phi} \, \bar{\Phi} \, ,
\label{eq:hatPhi}
 \\[1ex]
 \hat{q}^{a(\alpha)} =  
 \sqrt{Z_q} \, 
 q^{a(\alpha)} \, ,
&&~~
 \hat{\bar{q}}_{a(\alpha)} =  
 \sqrt{Z_q} \,
 \bar{q}_{a(\alpha)} \, .
\label{eq:hatQ}
\eeq
The theory exits from the conformal regime at the scale $M_c \sim \lambda \langle|\hat{\Phi}|\rangle \sim \bar{\lambda} \langle|\hat{\bar{\Phi}}|\rangle$.
Then, the $U(1)_{\rm PQ}$ breaking scale is given by 
\beq
f_{\rm PQ} &\equiv& 
\sqrt{\langle\hat{\Phi} \hat{\bar{\Phi}}\rangle} \simeq \frac{M_c}{\sqrt{\lambda\bar{\lambda}}}\nonumber\\
&=& \lmk\frac{\Lambda}{\Lambda_{\rm mag}}\rmk^{\frac{N_F(2N_F-3N)}{3(2N-N_F)(N_F-N)}} \frac{\Lambda}{\sqrt{\lambda\bar{\lambda}}} \lmk \frac{2M_\Phi}{N_F\lambda\bar{\lambda}\Lambda} \rmk^{\frac{N_F}{3(2N-N_F)}}\nonumber\\
&\simeq& \lmk\frac{\Lambda_{\rm mag}}{\Lambda}\rmk^{\frac{NN_F}{3(N_F-N)(2N-N_F)}} \frac{\Lambda}{\sqrt{\lambda\bar{\lambda}}} \lmk\frac{\iota N_F\Lambda}{2\lambda\bar{\lambda}\Mpl} \rmk^{\frac{2N-\delta}{3\delta}}.\nonumber\\
\label{PQscale}
\eeq
The third equality utilizes \EQ{MPhi} with $\Lambda'\simeq \Lambda_{\rm mag}$.
While the conformal entering scale $\Lambda$ is determined by solving the renormalization group (RG) equations
for the $SU(\widetilde{N})$ gauge coupling $g$ and $\lambda, \bar{\lambda}$ with initial conditions at a UV scale, 
$\Lambda_{\rm mag}$ is defined as the holomorphic dynamical scale in the magnetic picture. 
These scales are closely related to each other, depending on the initial condition of the gauge coupling at a UV scale
for RG equations. Without any fine tuning, we find $\Lambda/\Lambda_{\rm mag} = \mathcal{O}(1-10^2)$.
One can see from Eq.~\eqref{PQscale} that an intermediate scale of the $U(1)_{\rm PQ}$ breaking $f_{\rm PQ}$
is generated as the scale $\Lambda$ is naturally smaller than the Planck scale. 
\FIG{fig:hierarchy} summarizes hierarchies of scales in our model.
We take $M\simeq\Lambda$, so that extra singlet fields decouple at the conformal entering scale $\Lambda$.
In the following discussion, we focus on this case for simplicity.

We expand the scalar components of the PQ breaking fields around their VEVs,
$v\equiv |\langle\hat{\Phi}\rangle|$ and $\bar{v}\equiv |\langle\hat{\bar{\Phi}}\rangle|$:
\beq
\hat{\Phi} = v\exp\lmk \frac{\sigma +ia}{F_a}\rmk, \quad 
\hat{\bar{\Phi}} = \bar{v}\exp\lmk -\frac{\sigma +ia}{F_a}\rmk .
\eeq
Here, $\sigma, a$ represent the saxion and axion fields, respectively, and $F_a/\sqrt{2}\equiv\sqrt{v^2+\bar{v}^2}$.
After integrating out the heavy quark superfields $q,\bar{q}$, the effective theory has the axion-gluon coupling as well as the axion-photon coupling due to anomaly,
\beq
\Lag_{a} = \frac{\widetilde{N}}{2}\frac{g_s^2}{16\pi^2}\frac{a}{F_a}G_{\mu\nu}^i\widetilde{G}^{i\mu\nu}+E\frac{e^2}{16\pi^2}\frac{a}{F_a}F_{\mu\nu}\widetilde{F}^{\mu\nu} \, ,
\eeq
where $g_s, e$ are the gauge couplings for the $SU(3)_c$ and $U(1)_{\rm EM}$, respectively,
and $G,\widetilde{G}$ ($F,\widetilde{F}$) denote the gluon (photon) field strength and its dual.
The electromagnetic anomaly factor is given by $E = -n_f\widetilde{N} q_{\rm EM}^2$
where $q_{\rm EM}$ is the $U(1)_{\rm EM}$ charge of $q^a$ and $n_f$ denotes the number of $q^a$ with nonzero $U(1)_{\rm EM}$ charge.
Then, the axion decay constant can be defined as $F_a/\widetilde{N}$ with the domain wall number identified with $\widetilde{N}$.
Taking the basis without the gluon operator, we obtain the usual form of the axion-photon coupling, given by 
\beq
\Lag_{a\gamma} = \frac{g_{a\gamma}}{4}aF_{\mu\nu}\widetilde{F}^{\mu\nu} \, ,
\eeq
with
\beq
g_{a\gamma} =\frac{\alpha}{2\pi F_a/\widetilde{N}}\lmk -2n_fq_{\rm EM}^2-2.03\rmk.
\eeq
Here, $\alpha$ is the fine structure constant, and we have used $m_u/m_d=0.462$ \cite{ParticleDataGroup:2022pth}.
Since the sign of the first term in the parenthesis can be changed by the definition of the PQ charge, the axion coupling to photons can be enhanced or suppressed, depending on the $U(1)_Y$ charge assignment.

\begin{figure*}[t!]
    \begin{minipage}[t]{16cm}
    \includegraphics[width=7.5cm]{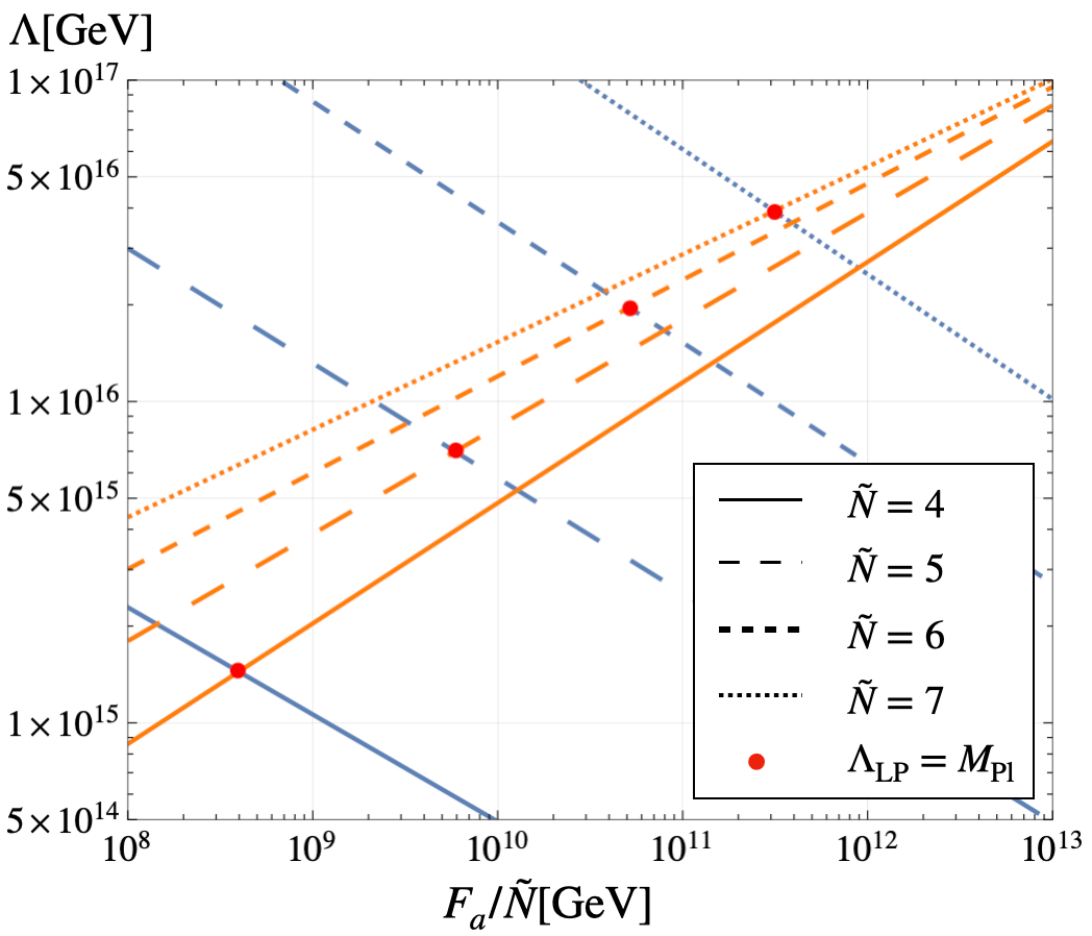}
    \hspace{8mm}
    \includegraphics[width=7.5cm]{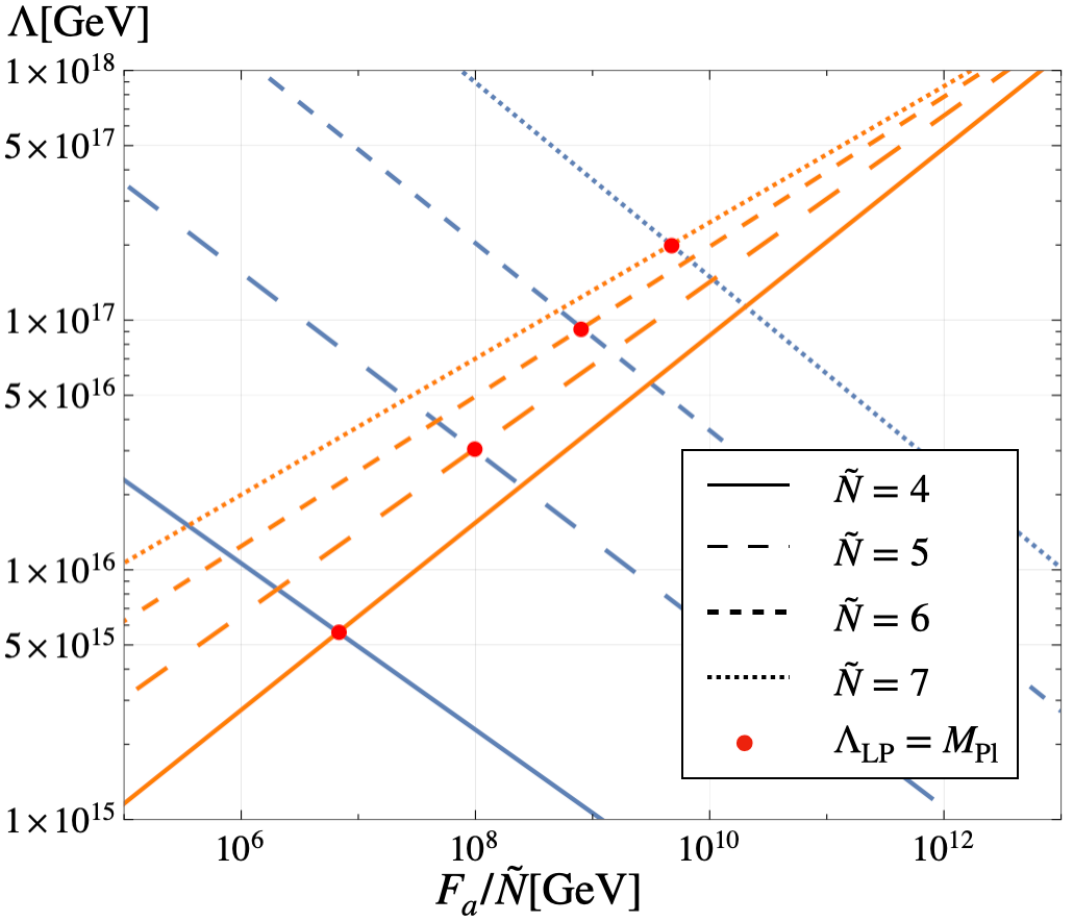}
    \end{minipage}
    \caption{
    The relation between the conformal entering scale $\Lambda$ and the axion decay constant $F_a/\widetilde{N}$ (orange lines).
    The blue line corresponds to the lower bound to solve the Landau pole problem
    where $\Lambda$ and $F_a/\widetilde{N}$ are assumed to be independent.
    The allowed region is above each blue line.
    The intersection of the orange and blue lines, represented by the red dot,
    is the actual lower bound on $\Lambda$ for each $N$. 
    Here, we take $N_F=2N-2,\iota=1, \Lambda/\Lambda_{\rm mag}=5 \, (100)$ in the left (right) panel, and use the values of $\lambda,\bar{\lambda}$ estimated at the fixed point.}
    \label{Landau_pole_para}
\end{figure*}

By using \EQ{PQscale}, the axion decay constant $F_a/\widetilde{N}$ is shown in \FIG{Landau_pole_para} (orange lines)
as a function of the conformal entering scale $\Lambda$. 
Here we take $N_F=2N-2$, $\iota=1$, and $\Lambda/\Lambda_{\rm mag}=5, 100$ in the left and right panels, respectively.
The values of $\lambda,\bar{\lambda}$ for each $\widetilde{N}$ are determined at the IR fixed point
and estimated at the two-loop level by using SARAH
\cite{Staub:2008uz} where QCD effects are ignored.
The figure demonstrates that an intermediate scale for $F_a/\widetilde{N}$ emerges
in the theory with much higher scales.

\subsection{SUSY breaking
\label{sec:saxion}}

There exists a flat direction in \EQ{flatdir}, which we call the saxion. 
In the SUSY limit, the saxion is massless, but SUSY breaking can deform the flat direction and induce a nonzero saxion mass.
Let us consider the following soft SUSY breaking terms for $\hat{\Phi}$, $\hat{\bar{\Phi}}$ and $\hat{q},\hat{\bar{q}}$
in the magnetic theory:
\beq
\mathcal{L}_{\rm soft} = 
&-&\hat{\tilde{q}}^{a*}m^{2}_{q^{a}}\hat{\tilde{q}}^{a} -\hat{\tilde{\bar{q}}}_{a}m^{2}_{\bar{q}_{a}}\hat{\tilde{\bar{q}}}^{*}_{a} \nonumber\\
&-&\hat{\tilde{q}}^{\alpha*}m^{2}_{q^{\alpha}}\hat{\tilde{q}}^{\alpha} -\hat{\tilde{\bar{q}}}_{\alpha}m^{2}_{\bar{q}_{\alpha}}\hat{\tilde{\bar{q}}}^{*}_{\alpha} 
\nonumber\\
&-& m^2_{\Phi}|\hat{\Phi}|^2
-m^2_{\bar{\Phi}}|\hat{\bar{\Phi}}|^2,
\label{soft}
\eeq
where $q,\bar{q}$ with tilde represent their scalar components,
and we have ignored SUSY breaking $A$-terms corresponding to the superpotential \eqref{integrate} for simplicity. 
All soft mass parameters are assumed at around the TeV scale, which
is much smaller than the $U(1)_{\rm PQ}$ breaking scale.
Taking $m_{q^a}^2 =  m_{\bar{q}_a}^2 = m_{q^{\alpha}}^2 = m_{\bar{q}_{\alpha}}^2 \equiv\Delta m^2$
and using Eq.~\eqref{integrate}, we obtain the one-loop Coleman-Weinberg potential \cite{Coleman:1973jx},
\beq   
&&V_{\rm CW}(\hat{\Phi},\hat{\bar{\Phi}})  \nonumber \\[1ex]
&&= 
\frac{\widetilde{N} N_F}{64\pi^2} 
\Biggl[\lmk\lambda^2|\hat{\Phi}|^2 + \Delta m^2\rmk^2 
\log\lmk 1+\frac{\Delta m^2}{\lambda^2|\hat{\Phi}|^2}\rmk \nonumber \\
&&\quad + \lmk\bar{\lambda}^2|\hat{\bar{\Phi}}|^2+\Delta m^2\rmk^2 
\log\lmk 1+\frac{\Delta m^2}{\bar{\lambda}^2|\hat{\bar{\Phi}}|^2}\rmk\Biggr] \, .
\label{CW}
\eeq
Summing up all contributions, the total scalar potential for $\Phi, \bar{\Phi}$ is expressed as
\beq
&&V(\hat{\Phi}, \hat{\bar{\Phi}}) = V_{\mathcal{F}} + V_{\rm soft} + V_{\rm CW} \, ,
\eeq
where Eq.~\eqref{VF} gives the first term of the right hand side and the soft mass-squared terms give the second contribution.

Since the $F$-term potential $V_{\mathcal{F}}$ is dominant, 
we can take $\hat{\Phi}\hat{\bar{\Phi}}=f_{\rm PQ}^2$ and focus on the flat direction. 
The VEVs for $\hat{\Phi}, \hat{\bar{\Phi}}$ with SUSY breaking effects are then obtained as
\beq
v &\equiv& \langle |\hat{\Phi}| \rangle \nonumber\\
&\simeq& f_{\rm PQ}\lmk\frac{(3\widetilde{N} N_F \lambda^2/64\pi^2)\Delta m^2 + N_Fm_{\Phi}^2/2}{(3\widetilde{N} N_F \bar{\lambda}^2/64\pi^2)\Delta m^2 + N_Fm_{\bar{\Phi}}^2/2}\rmk^{1/4} \!\!\! ,~~~\label{VEV}\\
\bar{v} &\equiv& \langle |\hat{\bar{\Phi}}| \rangle \nonumber\\
&\simeq& f_{\rm PQ} \lmk\frac{(3\widetilde{N} N_F \bar{\lambda}^2/64\pi^2)\Delta m^2 + N_Fm_{\bar{\Phi}}^2/2}{(3\widetilde{N} N_F \lambda^2/64\pi^2)\Delta m^2 + N_Fm_{\Phi}^2/2}\rmk^{1/4} \!\!\! .~~~
\label{VEV2}
\eeq
For $m_\Phi^2 = m_{\bar{\Phi}}^2$ and $\lambda = \bar{\lambda}$ which lead to $v=\bar{v}=f_{\rm PQ}$ and the approximation $\Delta m^2\ll f_{\rm PQ}^2$,
we can simply write the saxion mass-squared as 
\beq
m_\sigma^2 
\simeq  2m_\Phi^2+\frac{\widetilde{N}N_F\lambda^2}{32\pi^2}\Delta m^2 \, .
\eeq
Note that 
the saxion direction is stabilized only when the following condition is satisfied: 
\beq
m_\Phi^2 + \frac{\widetilde{N} N_F \lambda^2}{64\pi^2}\Delta m^2 > 0 \, .
\label{condition}
\eeq
This condition puts a constraint on the soft SUSY breaking parameters at the $U(1)_{\rm PQ}$ breaking scale $f_{\rm PQ}$.
These parameters evolve down to the scale $f_{\rm PQ}$ by RG equations, and hence
the proper initial values at a UV scale are required to satisfy the condition. 
For the RG evolution, see Fig.~2 in Ref.~\cite{Nakagawa:2023shi},
{which we have confirmed that is the same behavior with our case by replacing $N$ by $\widetilde{N}$}.

\section{Landau pole problem
\label{sec:Landau}}  

We now investigate the issue of a Landau pole for the SM $SU(3)_c$ gauge coupling $g_s$ in our model.
The one-loop beta function coefficient is given by
\begin{align}
\beta_{g_s}&\equiv\mu\frac{\dd g_s}{\dd\mu}\notag\\
&=-\frac{g_s^3}{16\pi^2}\times
\begin{cases}
    b_{\rm SM}\,, & M_{\rm EW}<\mu<M_{\rm SUSY}\, ,\\[1ex]
    b_{\rm MSSM}\,, & M_{\rm SUSY}<\mu<f_{\rm PQ}\, ,\\[1ex]
    b_{\rm mSQCD}\,, & f_{\rm PQ}<\mu<\Lambda\, ,\\[1ex]
    b_{\rm eSQCD}\,, & \Lambda<\mu<M_{\rm Pl}\, ,
\end{cases}\label{beta-func}
\end{align}
where $b_i~(i=\rm SM, \, MSSM, \, mSQCD, \, eSQCD)$ are calculated in different effective theories:
the SM, MSSM, magnetic picture and electric picture of our model, respectively.
The first two coefficients are given by $b_{\rm SM}=7$ and $b_{\rm MSSM}=3$ as usual.
Since the $SU(3)_c$ is embedded in the weakly gauged $SU(N_F/2)_1$,
the fields charged under $SU(N_F/2)_1$ in Tab.~\ref{tab:charge},\ref{tab:mSQCD charge}
contribute to $b_{\rm mSQCD}$ and $b_{\rm eSQCD}$. 
For the energy region of $f_{\rm PQ}<\mu<\Lambda$ where the magnetic picture gives a better description,
the fields $\eta,\xi, \Psi,\bar{\Psi}, \Sigma, \Xi$, which have been integrated out, as well as the $SU(N_F/2)_1$ singlets $\Phi,\bar{\Phi},q^\alpha,\bar{q}_\alpha$ do not contribute to $b_{\rm mSQCD}$.
The first half of magnetic quarks $q^a,\bar{q}_a$,
which are $\widetilde{N}=N_F-N$ fundamental and anti-fundamental representations of $SU(3)_c$,
contribute to $b_{\rm mSQCD}$.
Then, including the contributions from the $SU(3)_c$ gauge supermultiplet and the MSSM quarks,
we find
\begin{align}
b_{\rm mSQCD} &= 3T({\bf Adj}) - \sum\limits_{\rm MSSM} T(\square) - \sum\limits_{\rm mSQCD} T(\square) \notag \\
&= 3 \times 3 - 12\times\frac{1}{2} - 2\widetilde{N}\times\frac{1}{2} \notag \\
&= 3 - (N_F - N)\, ,
\end{align}
with $T({\bf Adj})=3$ and $T(\square)=T(\bar{\square})=\frac{1}{2}$. 
For the energy region of $\Lambda<\mu<M_{\rm Pl}$ where the electric theory is valid,
similarly, the singlets $\xi,Q_\alpha,\bar{Q}^\alpha$ do not contribute to $b_{\rm eSQCD}$,
while the electric quarks $Q_a,\bar{Q}^a$ are $N$ (anti-)fundamentals of $SU(3)_c$,
the extra fields $Z,\bar{Z}$ are $N_F/2$ (anti-)fundamentals
and $\eta$ is composed of one adjoint representation and $(N_F/2-3)$ (anti-)fundamentals.
The last statement comes from the decomposition of the adjoint of $SU(N_F/2)_1$:
\beq
{\bf Adj_{N_F/2}}= {\bf Adj_{3}}+\left(\frac{N_F}{2}-3\right)\left(\square_{\bf 3}+\bar{\square}_{\bf 3}\right)+{\bf 1}\, .
\eeq
Thus we obtain
\begin{align}
b_{\rm eSQCD}&= 3 \times 3 - 12\times\frac{1}{2} - 2N\times\frac{1}{2} \notag\\
&-2\frac{N_F}{2}\times\frac{1}{2}-2\left(\frac{N_F}{2}-3\right)\times\frac{1}{2}-3\notag \\
&= 3 - (N_F + N)\, .
\end{align}
We solve Eq.~\eqref{beta-func} and see how $g_s$ runs with the energy scale $\mu$.
For our typical values of $N$ and $N_F$, $b_{\rm mSQCD}$ and $b_{\rm eSQCD}$ are negative
and there exists a Landau pole for $g_s$ at a high energy scale $\Lambda_{\rm LP}$.
The Landau pole problem occurs when $\Lambda_{\rm LP}$ is smaller than the Planck scale $M_{\rm Pl}$,
which indicates that the theory breaks down before a UV theory above $M_{\rm Pl}$ appears.

Equivalent to Eq.~\eqref{beta-func}, we have a simpler form for the calculation of the scale of a Landau pole,
\begin{align}
\frac{\dd}{\dd\ln\mu}&\alpha_s^{-1} = \frac{b_i}{2\pi}\,,
\end{align}
where $\alpha_s = \frac{g_s^2}{4\pi}$ and $i=\rm SM, \, MSSM, \, mSQCD, \, eSQCD$.
Then, the Landau pole appears at
\begin{align}
\Lambda_{\rm LP} = \Lambda \exp &\ \left[-\frac{2\pi}{b_{\rm eSQCD}}\left(\alpha^{-1}_s(m_Z) + \frac{b_{\rm SM}}{2\pi} \ln{\frac{M_{\rm SUSY}}{m_Z}}\right.\right. \notag\\
&\left.\left. + \frac{b_{\rm MSSM}}{2\pi} \ln{\frac{f_{\rm PQ}}{M_{\rm SUSY}}} + \frac{b_{\rm mSQCD}}{2\pi} \ln{\frac{\Lambda}{f_{\rm PQ}}}\right)\right]\, 
\end{align}
with $m_Z=91.1876\rm GeV$ and $\alpha_s(m_Z)=0.1179$ \cite{ParticleDataGroup:2022pth}.
By requiring $\Lambda_{\rm LP}$ to be larger than $M_{\rm Pl}$,
we obtain the allowed region for the conformal entering scale $\Lambda$
and the $U(1)_{\rm PQ}$ breaking scale $f_{\rm PQ}$. The result is shown in Fig.~\ref{Landau_pole_para}, where we transform $f_{\rm PQ}$ into the axion decay constant by $F_a/\widetilde{N}\simeq 2 f_{\rm PQ}/\widetilde{N}$ in the limit of $v=\bar{v}=f_{\rm PQ}$.
For the blue lines in Fig.~\ref{Landau_pole_para}, $\Lambda$ and $F_a/\widetilde{N}$ are taken to be independent
to give a tendency for solving the Landau pole problem.
Here we take 
$N_F = 2N-2$ and $N=6,7,8,9$ for each line. The allowed region for each $N$ is above the blue line.
It shows that to solve the Landau pole problem requires a larger $\Lambda$ and a larger $F_a/\widetilde{N}$ ($f_{\rm PQ}$).
In fact, $\Lambda$ and $F_a/\widetilde{N}$ ($f_{\rm PQ}$) are related by Eq.~\eqref{PQscale}
and the relationship is represented by the orange lines. Thus the
intersection of each pair of the blue and orange lines, shown as the red dot, represents the actual lower bound for $\Lambda$ and $F_a/{\widetilde{N}}$. 
For $\Lambda/\Lambda_{\rm mag}=100$, we can take a larger value of $\Lambda$ to obtain the intermediate PQ scale, which shortens the range of the electric theory.  
Thus, one can see from the right panel that the range of the decay constant to avoid the Landau pole problem is broader than that of $\Lambda/\Lambda_{\rm mag}=5$.

\section{Quality of PQ symmetry
\label{sec:quality}}  

Let us discuss the quality of the PQ symmetry in the presence of Planck-suppressed operators
explicitly breaking the PQ symmetry,
and investigate whether the high PQ quality is consistent with the avoidance of the Landau pole problem.
Respecting the anomaly-free discrete symmetry $Z_{\widetilde{N}}$,
we can write the most dangerous PQ violating operators as
\beq
W_{\cancel{\rm PQ}} &=& c_{\cancel{\rm PQ}} \frac{(\bar{Q}^aQ_a)^{\widetilde{N}}}{\Mpl^{2\widetilde{N}-3}} +\bar{c}_{\cancel{\rm PQ}} \frac{(\bar{Q}^\alpha Q_\alpha)^{\widetilde{N}}}{\Mpl^{2\widetilde{N}-3}}\nonumber\\
&\leftrightarrow& 
Z_\Phi^{-\widetilde{N}/2}\lmk \frac{\Lambda'}{\Mpl} \rmk^{\widetilde{N}}
\frac{c_{\cancel{\rm PQ}} \hat{\Phi}^{\widetilde{N}}+\bar{c}_{\cancel{\rm PQ}}\hat{\bar{\Phi}}^{\widetilde{N}}}{\Mpl^{\widetilde{N}-3}} \, .
\label{WPQV}
\eeq
Here, $c_{\cancel{\rm PQ}}, \bar{c}_{\cancel{\rm PQ}}$ denote coupling coefficients, and we have used the duality \eqref{MQQ} and the renormalization of \EQ{eq:hatPhi} in the second line.
Comparing the present result with that of the model discussed in Ref.~\cite{Nakagawa:2023shi},
the same PQ violating operator receives an extra suppression factor
$(\Lambda'/\Mpl)^{\tilde{N}}$ in our model.
With a constant superpotential term, $W = m_{3/2} \Mpl^2$ where $m_{3/2}$ denotes the gravitino mass,
the scalar potential in supergravity, $V\supset -3WW^*/\Mpl^2$, leads to 
\begin{equation}
\begin{split}
V_{\cancel{\rm PQ}} = \left(\frac{M_c}{\Lambda}\right)^{\frac{\widetilde{N}\gamma_{\Phi}}{2}} \lmk \frac{\Lambda'}{\Mpl} \rmk^{\widetilde{N}} \frac{\kappa_{\cancel{\rm PQ}}m_{3/2}v^{\widetilde{N}}}{\Mpl^{\widetilde{N}-3}}\cos\left(\widetilde{N}\frac{a}{F_a}+\varphi\right),
\label{PQbreaking}
\end{split}
\end{equation}
where
$\kappa_{\cancel{\rm PQ}}$ is defined as an $\mathcal{O}(1)$ constant and $\varphi$ represents a phase. 
For simplicity, we have assumed $v\simeq\bar{v}$.

The potential \eqref{PQbreaking} should be compared with the axion potential generated from the ordinary QCD effect,
\beq
V_{\rm QCD} = \chi_0 \left[1-\cos\left(\widetilde{N}\frac{a}{F_a}\right)\right],
\label{QCDpot}
\eeq
where $\chi_0=(75.5\MeV)^4$ is the topological susceptibility at the zero temperature \cite{GrillidiCortona:2015jxo}. 
To quantitatively estimate the quality of the $U(1)_{\rm PQ}$ symmetry, we then introduce a quality factor defined by
\beq
\mathcal{Q}&\equiv& \frac{\left|V_{\rm \cancel{\rm PQ}}\right|_{\rm max}}{\left|V_{\rm QCD}\right|_{\rm max}} \nonumber\\
&\simeq&\lmk\frac{\Lambda_{\rm mag}}{\Lambda}\rmk^{\frac{(N_F-N)(3N-N_F)}{2N-N_F}} 
\kappa_{\cancel{\rm PQ}}(\lambda\bar{\lambda})^{-\frac{N_F(N_F-N)}{2(2N-N_F)}}\nonumber\\
&\times& \lmk\frac{\iota N_F}{2}\rmk^{\frac{(N_F-N)^2}{2N-N_F}} \lmk\frac{\Lambda}{\Mpl}\rmk^{\frac{(N_F-N)(3N-N_F)}{2N-N_F}}\nonumber\\
&\times& \frac{m_{3/2}\Mpl^3}{\chi_0}\sin\varphi \, .
\eeq
Here, the subscript ``max" denotes the maximum height of a potential and we use $\Lambda'\simeq \Lambda_{\rm mag}$.
Without fine tuning in the phase, $\varphi = \mathcal{O}(1)$, 
the potential minimum of the axion will be shifted by a factor of $\mathcal{Q}$ from the CP-conserving minimum. 
Considering the experimental constraint \cite{Abel:2020pzs},
the factor $\mathcal{Q}$ must be smaller than $10^{-10}$ to address the strong CP problem.

\begin{figure*}[t]
\begin{minipage}[t]{16cm}
\includegraphics[width=7.5cm]{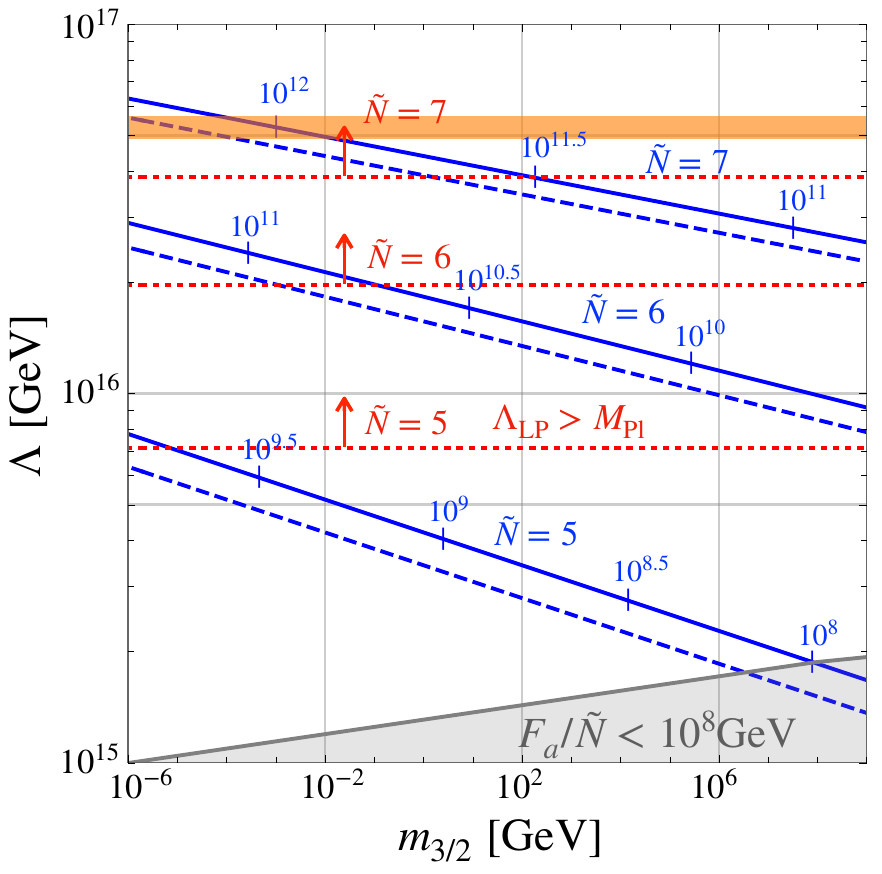}
\hspace{8mm}
\includegraphics[width=7.5cm]{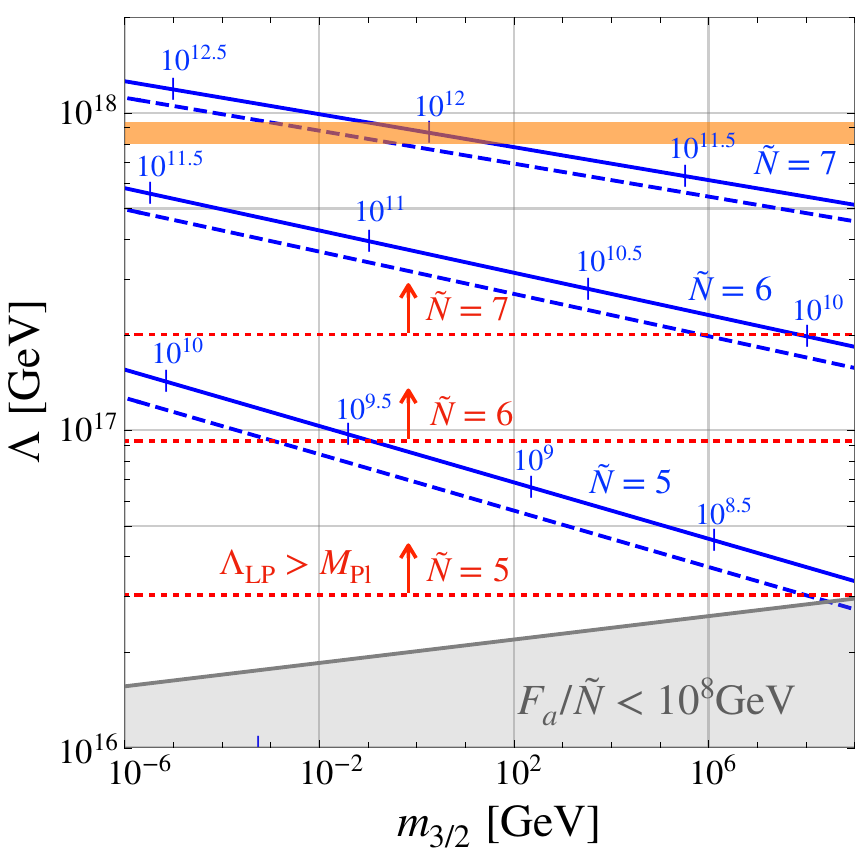}
\end{minipage}
\caption{
Contours of the quality factor for $\mathcal{Q}=10^{-10}$ (blue solid lines) and $\mathcal{Q}=10^{-12}$ (blue dashed lines) in the $m_{3/2}-\Lambda$ plane.
The scales on each blue line denote the decay constant $F_a/\tilde{N}$ in the unit of GeV. 
We take $\iota=1$, $\kappa_{\cancel{\rm PQ}}=1$, $N_F=2N-2$,
and $\varphi=1$, and utilize the values of $\lambda,\bar{\lambda}$ estimated at the IR fixed point.
The dynamical scale is taken as $\Lambda/\Lambda_{\rm mag}=5,100$ in the left and right panels, respectively.
The red dotted line represents a lower bound on $\Lambda$ from the Landau pole argument for each $\widetilde{N}\, (=N-2)$.
The gray shaded region is excluded by astrophysical bounds.
The orange band corresponds to the correct abundance of the axion DM, $\Omega_a \simeq \Omega_{\rm CDM}$.
}
\label{fig:quality}
\end{figure*}

\FIG{fig:quality} shows the contours for $\mathcal{Q}=10^{-10}$ (represented by the blue solid lines) and $\mathcal{Q}=10^{-12}$ (blue dashed lines) in the $m_{3/2}-\Lambda$ plane for $\widetilde{N}=5,6,7$ with $N_F=2N-2$, i.e. $\widetilde{N}=N-2$.
Here, we take $\Lambda/\Lambda_{\rm mag}=5~(100)$ in the left (right) panel and $\iota=\kappa_{\cancel{\rm PQ}}=\varphi=1$, and use the values of $\lambda,\bar{\lambda}$ estimated at the IR fixed point.
The scales on each line denote the axion decay constants $F_a/\tilde{N}$ in the unit of GeV.
The red dotted lines represent the constraint from the Landau pole problem, $\Lambda_{\rm LP}>\Mpl$,
and for each $N$, it puts a lower bound on $\Lambda$.
The gray shaded region corresponds to the astrophysical bounds, $F_a/\tilde{N}\lesssim 10^8\GeV$ \cite{Leinson:2014ioa,Hamaguchi:2018oqw,Leinson:2019cqv,Buschmann:2021juv,Mayle:1987as,Raffelt:1987yt,Turner:1987by,Chang:2018rso,Carenza:2019pxu}.
Below each blue solid line and above each red dotted line, the quality of $U(1)_{\rm PQ}$ is compatible with the absence of the Landau pole to a sufficiently good degree.
{In the range of $m_{3/2}$ shown here, the cases of $\widetilde{N}=6,7$ are promising for $\Lambda/\Lambda_{\rm mag}=5$.
On the other hand, one can see from the right panel that the quality of the PQ symmetry becomes much higher
for a larger $\Lambda/\Lambda_{\rm mag}$.
This is because the suppression due to the wavefunction renormalization becomes significant.
In this case, the theory with $\widetilde{N}\geq5$ has the high quality PQ symmetry without a Landau pole problem
in a broader range of the gravitino mass,
which can be compared with the previous work \cite{Nakai:2021nyf} where the gravitino mass is limited below $10\GeV$.
In addition, one can see that the $SU(\widetilde{N})$ gauge theory with $\widetilde{N}=5$ becomes viable
while it is not in the setup of Ref.~\cite{Nakagawa:2023shi}.}

Let us comment on the value of $\delta$.
Although we have mainly shown the results of $\delta=2$, a larger value is also possible, as long as $\widetilde{N}$ is sufficiently large for the theory to be in conformal window.
For example, $\widetilde{N}\geq7$ is required for $\delta=6$.
However, we have found that a larger value of $\delta$ does not improve the PQ quality and the Landau pole problem.

\section{DM axion implications
\label{sec:DM}}

Our axion model to address the strong CP problem has a viable parameter space consistent with the range of the axion decay constant, $10^{10}\GeV\lesssim F_a/\widetilde{N} \lesssim 10^{12}\GeV$, where the axion can give the observed abundance of DM.
At the QCD phase transition, the axion DM is produced via the misalignment mechanism \cite{Preskill:1982cy,Abbott:1982af,Dine:1982ah}.
When the PQ symmetry is spontaneously broken before inflation and is never restored after inflation,\footnote{If the PQ symmetry was spontaneously broken after inflation, topological defects would easily dominate the Universe,
because the domain wall number is larger than 1 in our model.}
the axion DM abundance is obtained as
\cite{Ballesteros:2016xej}
\beq
\Omega_a h^2 \simeq 0.14  \,
\theta_{\rm ini}^2\lmk\frac{F_a/\widetilde{N}}{10^{12}\GeV}\rmk^{1.17},
\eeq
where $\theta_{\rm ini} \equiv a_{\rm ini}/(F_a/\widetilde{N})$ is defined as
the dimensionless initial position of the axion field. 
Here we omit the anharmonic effect whose contribution is not significant for the mass range of interest.
The abundance can be matched with the observed DM abundance for $F_a/\widetilde{N}\simeq10^{12}\GeV$ and $\theta_{\rm ini}=\mathcal{O}(1)$.
The orange band in \FIG{fig:quality} shows the region with the correct abundance of DM $(F_a/\widetilde{N}\sim10^{12}\GeV)$,
which indicates $\widetilde{N}=7, N_F=16$,
and $m_{3/2}\lesssim 1\MeV~ (1\GeV)$ for $\Lambda/\Lambda_{\rm mag}=5~(100)$.
It is interesting to note that quantum gravity or some other UV effects can leave footprints
through the explicit PQ symmetry violation in this region, which will be probed by
future neutron EDM experiments, such as the TUCAN, nEDM, n2EDM \cite{TUCAN:2018vmr,nEDM:2019qgk,n2EDM:2021yah}.

In the other region of the parameter space, the axion is only a subdominant component of DM.
Since our theory is reduced to the MSSM at the electroweak scale, the lightest supersymmetric particle (LSP),
such as the neutralino, can become a viable DM candidate, with the conserved $R$-parity
\cite{Jungman:1995df}.
It is then important to note that our high-quality axion model can accommodate a heavy gravitino,
$m_{3/2} \gtrsim 10-100 \TeV$, where the mini-split-type SUSY spectrum
\cite{Ibe:2011aa,Ibe:2012hu,Arvanitaki:2012ps,Arkani-Hamed:2012fhg} can be realized.

Since the axion is very light and acquires quantum fluctuation during inflation,
the axionic isocurvature fluctuation imposes an upper bound on the Hubble parameter $H_{\rm inf}$ during inflation \cite{Steinhardt:1983ia,Axenides:1983hj,Linde:1985yf,Seckel:1985tj}.
The isocurvature fluctuation of cold dark matter (CDM) is given by \cite{Kobayashi:2013nva}
\beq
S_{\rm CDM} \simeq \frac{\delta\rho_{\rm CDM}}{\rho_{\rm CDM}} \simeq \frac{\Omega_a}{\Omega_{\rm CDM}}\frac{\delta\rho_a}{\rho_a}\simeq\frac{\Omega_a}{\Omega_{\rm CDM}}\frac{2\delta a_*}{a_*} \, ,
\eeq
where $\rho_{a(\rm CDM)},\delta\rho_{a(\rm CDM)}$  respectively represent the axion (CDM) energy density and its fluctuation,
$\Omega_{a({\rm CDM})}$ is the density parameter for the axion (CDM),
and $a_*,\delta a_*$ are the axion field value and its fluctuation during inflation. 
The third equality uses the fact that $\rho_a\propto a_*^2$.
Here the photon fluctuation is ignored with a good approximation,
and it is assumed that the isocurvature fluctuation is generated only from the axion.
The recent Planck data of anisotropy of cosmic microwave background radiation constrains the isocurvature fluctuation \cite{Planck:2018jri},
\beq
\beta_{\rm iso} (k_*) <0.038 \, ,
\eeq
where $\beta_{\rm iso}$ is defined as the ratio between the power spectrum of the adiabatic and isocurvature fluctuations
at the scale of $k_*=0.05 \, {\rm Mpc}^{-1}$.
For a natural initial misalignment angle which gives the correct abundance of DM,
the upper bound on the scale of inflation is given by
\beq
\label{Hinf}
H_{\rm inf} \lesssim 2.7\times 10^{7}\GeV \lmk\frac{F_a/\widetilde{N}}{10^{12}\GeV}\rmk^{0.42}.
\eeq
Compared to the current upper bound from the tensor-to-scalar ratio \cite{Planck:2018jri}, $H_{\rm inf}<6.5\times10^{13}\GeV$,
the upper bound \eqref{Hinf} gives a stringent constraint on inflation models.

A possible way out of the axion isocurvature problem is the Linde's solution
\cite{Linde:1991km}
where PQ breaking fields acquire large VEVs during inflation, and the axion fluctuation is significantly suppressed, $\delta \theta_*\simeq \widetilde{N}H_{\rm inf}/2\pi\Mpl$.
In this case, the upper bound on the Hubble parameter during inflation is given by 
\beq
H_{\rm inf}\lesssim 9.1\times 10^{12}\GeV \lmk\frac{7}{\widetilde{N}}\rmk \lmk\frac{10^{12}\GeV}{F_a/\widetilde{N}}\rmk^{0.59} \!\!\!,
\eeq
where we use $\Omega_a=\Omega_{\rm CDM}$.
This is comparable to the current upper bound from the tensor-to-scalar ratio. 
However, it has been recognized that this mechanism suffers from the enhancement of fluctuations of
the PQ breaking scalar fields produced by parametric resonance \cite{Kofman:1994rk}.
Due to those fluctuations, the PQ symmetry may be restored after inflation, which leads to the domain wall problem \cite{Kasuya:1996ns,Kawasaki:2013iha,Kawasaki:2017kkr,Kawasaki:2018qwp}.

\begin{figure}[t]
\centering
\includegraphics[width=6cm]{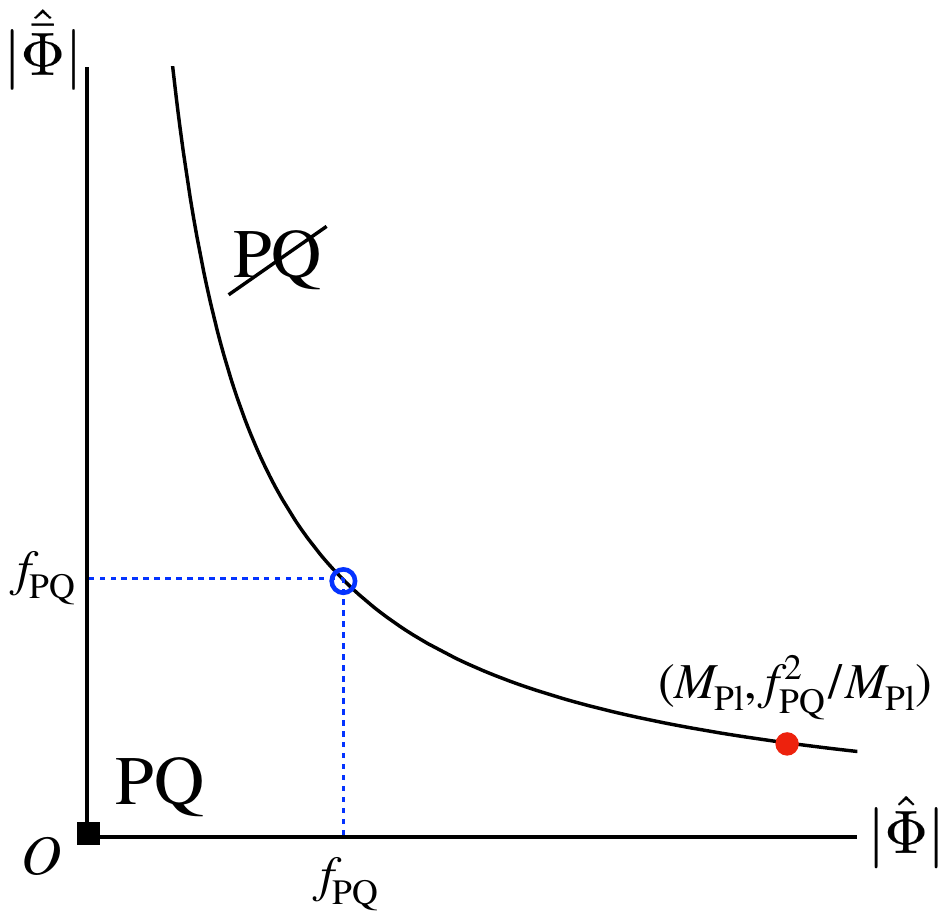}
\caption{The schematic picture of the minima of the PQ breaking fields $(|\hat{\Phi}|,|\hat{\bar{\Phi}}|)$.
The black square at the origin denotes the PQ symmetric point of \EQ{PQsymm},
and the black line gives the flat direction in \EQ{flatdir}.
The field $|\hat{\Phi}|$ obtains a large VEV (red bullet) during inflation due to the Hubble induced mass term.
When SUSY breaking becomes relevant after inflation, the oscillation along the flat direction starts around the point $(f_{\rm PQ},f_{\rm PQ})$ (blue circle).
}
\label{fig:flatdir}
\end{figure}

Refs.~\cite{Kasuya:1996ns,Kawasaki:2017kkr,Kawasaki:2018qwp} have discussed that
a SUSY axion model with a flat direction has a possibility to overcome the shortcoming of the Linde's solution.
The argument can be applied to our axion model with \EQ{flatdir}
whose flat direction is deformed by the Hubble induced mass.
The scalar potential of $\Phi,\bar{\Phi}$ is then given by
\beq
V = V_{\mathcal{F}} + V_{\rm soft} + V_{\rm CW} + cH^2|\Phi|^2 + \bar{c}H^2|\bar{\Phi}|^2 \, ,
\label{Hubble}
\eeq
where $H$ denotes the Hubble parameter, and $c,\bar{c}$ are dimensionless constants. 
The soft SUSY breaking terms do not affect the dynamics during inflation but are relevant to the oscillation after inflation. 
When we take e.g. $c<0$ and $\bar{c}>0$, $\hat{\Phi},\hat{\bar{\Phi}}$ are expected
to acquire VEVs, $v\simeq \Mpl$ and $\bar{v}\simeq f_{\rm PQ}^2/\Mpl$, during inflation,
so that the mechanism to suppress the axion fluctuation works well in our model.
\FIG{fig:flatdir} shows the schematic picture of the vacuum in the $(|\hat{\Phi}|,|\hat{\bar{\Phi}}|)$ plane,
and the VEVs during inflation are denoted by the red bullet on the flat direction (black line). 
When the oscillation along the flat direction starts around the final vacuum (blue circle) due to SUSY breaking,
the fluctuations of the PQ breaking scalar fields are generated via parametric resonance.
However, those fluctuations never restore the PQ symmetry.
There is a potential issue that the fluctuations
make the distribution of the axion field completely flat, which may generate stable domain walls, but
it is unlikely because we can expect that a peak of the axion field fluctuation evolves with time
due to the axion gradient term. 
For a further confirmation, numerical simulations are needed
\cite{Kawasaki:2017kkr}.

\section{Discussions
\label{sec:Discussion}}

We have explored a new paradigm for the high-quality QCD axion
on the basis of electric-magnetic duality in the conformal window of a SUSY gauge theory.
The PQ breaking fields emerge in the magnetic picture of the theory. 
Their large anomalous dimension leads to not only the intermediate scale of the spontaneous PQ violation
but also a significant suppression of explicit PQ symmetry breaking operators.
The high PQ quality and the absence of a Landau pole in the color gauge coupling
are achieved at the same time.
The parameter space to realize the correct abundance of the axion DM with the decay constant $f_a=10^{10}-10^{12}\GeV$ predicts
explicit PQ violation which may be probed by future measurements of the neutron EDM.
The axion isocurvature fluctuation can be naturally suppressed by large VEVs of the PQ breaking fields during inflation.
In the other viable parameter space, the LSP can become a DM candidate.
The model accommodates a heavy gravitino,
$m_{3/2} \gtrsim 10-100 \TeV$, where the mini-split-type SUSY spectrum is realized.
Therefore, our high-quality axion model provides a complete solution to the strong CP problem as well as the identity of DM.

A supersymmetric axion model contains a scalar partner, the saxion, as well as a fermionic partner called the axino. 
In \SEC{sec:saxion}, we have discussed the stabilization of the saxion with SUSY breaking effects,
and it was found that its mass is given by the soft mass scale $\sim 1 \TeV$.
Such a saxion mode can easily decay into gluons so that any cosmological problem is not induced.
On the other hand, the axino can acquire a mass $m_{\widetilde{a}} = \mathcal{O}(m_{3/2})$
from supergravity effects, $\Lag\supset \int d^4\theta (A+A^\dagger)(X+X^\dagger)/\Mpl$,
where $A$ and $X$ respectively denote the axion and SUSY breaking superfields. 
Cosmologically, the overproduction of the axino puts an upper bound on the reheating temperature
in some range of $m_{3/2}$ \cite{Cheung:2011mg}.

\section*{Acknowledgments}
We thank Motoo Suzuki and Masaki Yamada for useful discussions.
YN is supported by Natural Science Foundation of Shanghai.

\appendix

\bibliography{reference}

\begin{thebibliography}{102}%
\makeatletter
\providecommand \@ifxundefined [1]{%
 \@ifx{#1\undefined}
}%
\providecommand \@ifnum [1]{%
 \ifnum #1\expandafter \@firstoftwo
 \else \expandafter \@secondoftwo
 \fi
}%
\providecommand \@ifx [1]{%
 \ifx #1\expandafter \@firstoftwo
 \else \expandafter \@secondoftwo
 \fi
}%
\providecommand \natexlab [1]{#1}%
\providecommand \enquote  [1]{``#1''}%
\providecommand \bibnamefont  [1]{#1}%
\providecommand \bibfnamefont [1]{#1}%
\providecommand \citenamefont [1]{#1}%
\providecommand \href@noop [0]{\@secondoftwo}%
\providecommand \href [0]{\begingroup \@sanitize@url \@href}%
\providecommand \@href[1]{\@@startlink{#1}\@@href}%
\providecommand \@@href[1]{\endgroup#1\@@endlink}%
\providecommand \@sanitize@url [0]{\catcode `\\12\catcode `\$12\catcode
  `\&12\catcode `\#12\catcode `\^12\catcode `\_12\catcode `\%12\relax}%
\providecommand \@@startlink[1]{}%
\providecommand \@@endlink[0]{}%
\providecommand \url  [0]{\begingroup\@sanitize@url \@url }%
\providecommand \@url [1]{\endgroup\@href {#1}{\urlprefix }}%
\providecommand \urlprefix  [0]{URL }%
\providecommand \Eprint [0]{\href }%
\providecommand \doibase [0]{http://dx.doi.org/}%
\providecommand \selectlanguage [0]{\@gobble}%
\providecommand \bibinfo  [0]{\@secondoftwo}%
\providecommand \bibfield  [0]{\@secondoftwo}%
\providecommand \translation [1]{[#1]}%
\providecommand \BibitemOpen [0]{}%
\providecommand \bibitemStop [0]{}%
\providecommand \bibitemNoStop [0]{.\EOS\space}%
\providecommand \EOS [0]{\spacefactor3000\relax}%
\providecommand \BibitemShut  [1]{\csname bibitem#1\endcsname}%
\let\auto@bib@innerbib\@empty
\bibitem [{\citenamefont {Baker}\ \emph {et~al.}(2006)\citenamefont {Baker}
  \emph {et~al.}}]{Baker:2006ts}%
  \BibitemOpen
  \bibfield  {author} {\bibinfo {author} {\bibfnamefont {C.~A.}\ \bibnamefont
  {Baker}} \emph {et~al.},\ }\href {\doibase 10.1103/PhysRevLett.97.131801}
  {\bibfield  {journal} {\bibinfo  {journal} {Phys. Rev. Lett.}\ }\textbf
  {\bibinfo {volume} {97}},\ \bibinfo {pages} {131801} (\bibinfo {year}
  {2006})},\ \Eprint {http://arxiv.org/abs/hep-ex/0602020}
  {arXiv:hep-ex/0602020} \BibitemShut {NoStop}%
\bibitem [{\citenamefont {Pendlebury}\ \emph {et~al.}(2015)\citenamefont
  {Pendlebury} \emph {et~al.}}]{Pendlebury:2015lrz}%
  \BibitemOpen
  \bibfield  {author} {\bibinfo {author} {\bibfnamefont {J.~M.}\ \bibnamefont
  {Pendlebury}} \emph {et~al.},\ }\href {\doibase 10.1103/PhysRevD.92.092003}
  {\bibfield  {journal} {\bibinfo  {journal} {Phys. Rev. D}\ }\textbf {\bibinfo
  {volume} {92}},\ \bibinfo {pages} {092003} (\bibinfo {year} {2015})},\
  \Eprint {http://arxiv.org/abs/1509.04411} {arXiv:1509.04411 [hep-ex]}
  \BibitemShut {NoStop}%
\bibitem [{\citenamefont {Peccei}\ and\ \citenamefont
  {Quinn}(1977)}]{Peccei:1977hh}%
  \BibitemOpen
  \bibfield  {author} {\bibinfo {author} {\bibfnamefont {R.~D.}\ \bibnamefont
  {Peccei}}\ and\ \bibinfo {author} {\bibfnamefont {H.~R.}\ \bibnamefont
  {Quinn}},\ }\href {\doibase 10.1103/PhysRevLett.38.1440} {\bibfield
  {journal} {\bibinfo  {journal} {Phys. Rev. Lett.}\ }\textbf {\bibinfo
  {volume} {38}},\ \bibinfo {pages} {1440} (\bibinfo {year}
  {1977})}\BibitemShut {NoStop}%
\bibitem [{\citenamefont {Weinberg}(1978)}]{Weinberg:1977ma}%
  \BibitemOpen
  \bibfield  {author} {\bibinfo {author} {\bibfnamefont {S.}~\bibnamefont
  {Weinberg}},\ }\href {\doibase 10.1103/PhysRevLett.40.223} {\bibfield
  {journal} {\bibinfo  {journal} {Phys. Rev. Lett.}\ }\textbf {\bibinfo
  {volume} {40}},\ \bibinfo {pages} {223} (\bibinfo {year} {1978})}\BibitemShut
  {NoStop}%
\bibitem [{\citenamefont {Wilczek}(1978)}]{Wilczek:1977pj}%
  \BibitemOpen
  \bibfield  {author} {\bibinfo {author} {\bibfnamefont {F.}~\bibnamefont
  {Wilczek}},\ }\href {\doibase 10.1103/PhysRevLett.40.279} {\bibfield
  {journal} {\bibinfo  {journal} {Phys. Rev. Lett.}\ }\textbf {\bibinfo
  {volume} {40}},\ \bibinfo {pages} {279} (\bibinfo {year} {1978})}\BibitemShut
  {NoStop}%
\bibitem [{\citenamefont {Preskill}\ \emph {et~al.}(1983)\citenamefont
  {Preskill}, \citenamefont {Wise},\ and\ \citenamefont
  {Wilczek}}]{Preskill:1982cy}%
  \BibitemOpen
  \bibfield  {author} {\bibinfo {author} {\bibfnamefont {J.}~\bibnamefont
  {Preskill}}, \bibinfo {author} {\bibfnamefont {M.~B.}\ \bibnamefont {Wise}},
  \ and\ \bibinfo {author} {\bibfnamefont {F.}~\bibnamefont {Wilczek}},\ }\href
  {\doibase 10.1016/0370-2693(83)90637-8} {\bibfield  {journal} {\bibinfo
  {journal} {Phys. Lett. B}\ }\textbf {\bibinfo {volume} {120}},\ \bibinfo
  {pages} {127} (\bibinfo {year} {1983})}\BibitemShut {NoStop}%
\bibitem [{\citenamefont {Abbott}\ and\ \citenamefont
  {Sikivie}(1983)}]{Abbott:1982af}%
  \BibitemOpen
  \bibfield  {author} {\bibinfo {author} {\bibfnamefont {L.~F.}\ \bibnamefont
  {Abbott}}\ and\ \bibinfo {author} {\bibfnamefont {P.}~\bibnamefont
  {Sikivie}},\ }\href {\doibase 10.1016/0370-2693(83)90638-X} {\bibfield
  {journal} {\bibinfo  {journal} {Phys. Lett. B}\ }\textbf {\bibinfo {volume}
  {120}},\ \bibinfo {pages} {133} (\bibinfo {year} {1983})}\BibitemShut
  {NoStop}%
\bibitem [{\citenamefont {Dine}\ and\ \citenamefont
  {Fischler}(1983)}]{Dine:1982ah}%
  \BibitemOpen
  \bibfield  {author} {\bibinfo {author} {\bibfnamefont {M.}~\bibnamefont
  {Dine}}\ and\ \bibinfo {author} {\bibfnamefont {W.}~\bibnamefont
  {Fischler}},\ }\href {\doibase 10.1016/0370-2693(83)90639-1} {\bibfield
  {journal} {\bibinfo  {journal} {Phys. Lett. B}\ }\textbf {\bibinfo {volume}
  {120}},\ \bibinfo {pages} {137} (\bibinfo {year} {1983})}\BibitemShut
  {NoStop}%
\bibitem [{\citenamefont {Mayle}\ \emph {et~al.}(1988)\citenamefont {Mayle},
  \citenamefont {Wilson}, \citenamefont {Ellis}, \citenamefont {Olive},
  \citenamefont {Schramm},\ and\ \citenamefont {Steigman}}]{Mayle:1987as}%
  \BibitemOpen
  \bibfield  {author} {\bibinfo {author} {\bibfnamefont {R.}~\bibnamefont
  {Mayle}}, \bibinfo {author} {\bibfnamefont {J.~R.}\ \bibnamefont {Wilson}},
  \bibinfo {author} {\bibfnamefont {J.~R.}\ \bibnamefont {Ellis}}, \bibinfo
  {author} {\bibfnamefont {K.~A.}\ \bibnamefont {Olive}}, \bibinfo {author}
  {\bibfnamefont {D.~N.}\ \bibnamefont {Schramm}}, \ and\ \bibinfo {author}
  {\bibfnamefont {G.}~\bibnamefont {Steigman}},\ }\href {\doibase
  10.1016/0370-2693(88)91595-X} {\bibfield  {journal} {\bibinfo  {journal}
  {Phys. Lett. B}\ }\textbf {\bibinfo {volume} {203}},\ \bibinfo {pages} {188}
  (\bibinfo {year} {1988})}\BibitemShut {NoStop}%
\bibitem [{\citenamefont {Raffelt}\ and\ \citenamefont
  {Seckel}(1988)}]{Raffelt:1987yt}%
  \BibitemOpen
  \bibfield  {author} {\bibinfo {author} {\bibfnamefont {G.}~\bibnamefont
  {Raffelt}}\ and\ \bibinfo {author} {\bibfnamefont {D.}~\bibnamefont
  {Seckel}},\ }\href {\doibase 10.1103/PhysRevLett.60.1793} {\bibfield
  {journal} {\bibinfo  {journal} {Phys. Rev. Lett.}\ }\textbf {\bibinfo
  {volume} {60}},\ \bibinfo {pages} {1793} (\bibinfo {year}
  {1988})}\BibitemShut {NoStop}%
\bibitem [{\citenamefont {Turner}(1988)}]{Turner:1987by}%
  \BibitemOpen
  \bibfield  {author} {\bibinfo {author} {\bibfnamefont {M.~S.}\ \bibnamefont
  {Turner}},\ }\href {\doibase 10.1103/PhysRevLett.60.1797} {\bibfield
  {journal} {\bibinfo  {journal} {Phys. Rev. Lett.}\ }\textbf {\bibinfo
  {volume} {60}},\ \bibinfo {pages} {1797} (\bibinfo {year}
  {1988})}\BibitemShut {NoStop}%
\bibitem [{\citenamefont {Chang}\ \emph {et~al.}(2018)\citenamefont {Chang},
  \citenamefont {Essig},\ and\ \citenamefont {McDermott}}]{Chang:2018rso}%
  \BibitemOpen
  \bibfield  {author} {\bibinfo {author} {\bibfnamefont {J.~H.}\ \bibnamefont
  {Chang}}, \bibinfo {author} {\bibfnamefont {R.}~\bibnamefont {Essig}}, \ and\
  \bibinfo {author} {\bibfnamefont {S.~D.}\ \bibnamefont {McDermott}},\ }\href
  {\doibase 10.1007/JHEP09(2018)051} {\bibfield  {journal} {\bibinfo  {journal}
  {JHEP}\ }\textbf {\bibinfo {volume} {09}},\ \bibinfo {pages} {051} (\bibinfo
  {year} {2018})},\ \Eprint {http://arxiv.org/abs/1803.00993} {arXiv:1803.00993
  [hep-ph]} \BibitemShut {NoStop}%
\bibitem [{\citenamefont {Carenza}\ \emph {et~al.}(2019)\citenamefont
  {Carenza}, \citenamefont {Fischer}, \citenamefont {Giannotti}, \citenamefont
  {Guo}, \citenamefont {Mart\'\i{}nez-Pinedo},\ and\ \citenamefont
  {Mirizzi}}]{Carenza:2019pxu}%
  \BibitemOpen
  \bibfield  {author} {\bibinfo {author} {\bibfnamefont {P.}~\bibnamefont
  {Carenza}}, \bibinfo {author} {\bibfnamefont {T.}~\bibnamefont {Fischer}},
  \bibinfo {author} {\bibfnamefont {M.}~\bibnamefont {Giannotti}}, \bibinfo
  {author} {\bibfnamefont {G.}~\bibnamefont {Guo}}, \bibinfo {author}
  {\bibfnamefont {G.}~\bibnamefont {Mart\'\i{}nez-Pinedo}}, \ and\ \bibinfo
  {author} {\bibfnamefont {A.}~\bibnamefont {Mirizzi}},\ }\href {\doibase
  10.1088/1475-7516/2019/10/016} {\bibfield  {journal} {\bibinfo  {journal}
  {JCAP}\ }\textbf {\bibinfo {volume} {10}},\ \bibinfo {pages} {016} (\bibinfo
  {year} {2019})},\ \bibinfo {note} {[Erratum: JCAP 05, E01 (2020)]},\ \Eprint
  {http://arxiv.org/abs/1906.11844} {arXiv:1906.11844 [hep-ph]} \BibitemShut
  {NoStop}%
\bibitem [{\citenamefont {Leinson}(2014)}]{Leinson:2014ioa}%
  \BibitemOpen
  \bibfield  {author} {\bibinfo {author} {\bibfnamefont {L.~B.}\ \bibnamefont
  {Leinson}},\ }\href {\doibase 10.1088/1475-7516/2014/08/031} {\bibfield
  {journal} {\bibinfo  {journal} {JCAP}\ }\textbf {\bibinfo {volume} {08}},\
  \bibinfo {pages} {031} (\bibinfo {year} {2014})},\ \Eprint
  {http://arxiv.org/abs/1405.6873} {arXiv:1405.6873 [hep-ph]} \BibitemShut
  {NoStop}%
\bibitem [{\citenamefont {Hamaguchi}\ \emph {et~al.}(2018)\citenamefont
  {Hamaguchi}, \citenamefont {Nagata}, \citenamefont {Yanagi},\ and\
  \citenamefont {Zheng}}]{Hamaguchi:2018oqw}%
  \BibitemOpen
  \bibfield  {author} {\bibinfo {author} {\bibfnamefont {K.}~\bibnamefont
  {Hamaguchi}}, \bibinfo {author} {\bibfnamefont {N.}~\bibnamefont {Nagata}},
  \bibinfo {author} {\bibfnamefont {K.}~\bibnamefont {Yanagi}}, \ and\ \bibinfo
  {author} {\bibfnamefont {J.}~\bibnamefont {Zheng}},\ }\href {\doibase
  10.1103/PhysRevD.98.103015} {\bibfield  {journal} {\bibinfo  {journal} {Phys.
  Rev. D}\ }\textbf {\bibinfo {volume} {98}},\ \bibinfo {pages} {103015}
  (\bibinfo {year} {2018})},\ \Eprint {http://arxiv.org/abs/1806.07151}
  {arXiv:1806.07151 [hep-ph]} \BibitemShut {NoStop}%
\bibitem [{\citenamefont {Leinson}(2019)}]{Leinson:2019cqv}%
  \BibitemOpen
  \bibfield  {author} {\bibinfo {author} {\bibfnamefont {L.~B.}\ \bibnamefont
  {Leinson}},\ }\href {\doibase 10.1088/1475-7516/2019/11/031} {\bibfield
  {journal} {\bibinfo  {journal} {JCAP}\ }\textbf {\bibinfo {volume} {11}},\
  \bibinfo {pages} {031} (\bibinfo {year} {2019})},\ \Eprint
  {http://arxiv.org/abs/1909.03941} {arXiv:1909.03941 [hep-ph]} \BibitemShut
  {NoStop}%
\bibitem [{\citenamefont {Buschmann}\ \emph {et~al.}(2022)\citenamefont
  {Buschmann}, \citenamefont {Dessert}, \citenamefont {Foster}, \citenamefont
  {Long},\ and\ \citenamefont {Safdi}}]{Buschmann:2021juv}%
  \BibitemOpen
  \bibfield  {author} {\bibinfo {author} {\bibfnamefont {M.}~\bibnamefont
  {Buschmann}}, \bibinfo {author} {\bibfnamefont {C.}~\bibnamefont {Dessert}},
  \bibinfo {author} {\bibfnamefont {J.~W.}\ \bibnamefont {Foster}}, \bibinfo
  {author} {\bibfnamefont {A.~J.}\ \bibnamefont {Long}}, \ and\ \bibinfo
  {author} {\bibfnamefont {B.~R.}\ \bibnamefont {Safdi}},\ }\href {\doibase
  10.1103/PhysRevLett.128.091102} {\bibfield  {journal} {\bibinfo  {journal}
  {Phys. Rev. Lett.}\ }\textbf {\bibinfo {volume} {128}},\ \bibinfo {pages}
  {091102} (\bibinfo {year} {2022})},\ \Eprint
  {http://arxiv.org/abs/2111.09892} {arXiv:2111.09892 [hep-ph]} \BibitemShut
  {NoStop}%
\bibitem [{\citenamefont {O'Hare}(2020)}]{AxionLimits}%
  \BibitemOpen
  \bibfield  {author} {\bibinfo {author} {\bibfnamefont {C.}~\bibnamefont
  {O'Hare}},\ }\href {\doibase 10.5281/zenodo.3932430} {\enquote {\bibinfo
  {title} {cajohare/axionlimits: Axionlimits},}\ }\bibinfo {howpublished}
  {\url{https://cajohare.github.io/AxionLimits/}} (\bibinfo {year}
  {2020})\BibitemShut {NoStop}%
\bibitem [{\citenamefont {Dine}\ and\ \citenamefont
  {Seiberg}(1986)}]{Dine:1986bg}%
  \BibitemOpen
  \bibfield  {author} {\bibinfo {author} {\bibfnamefont {M.}~\bibnamefont
  {Dine}}\ and\ \bibinfo {author} {\bibfnamefont {N.}~\bibnamefont {Seiberg}},\
  }\href {\doibase 10.1016/0550-3213(86)90043-X} {\bibfield  {journal}
  {\bibinfo  {journal} {Nucl. Phys. B}\ }\textbf {\bibinfo {volume} {273}},\
  \bibinfo {pages} {109} (\bibinfo {year} {1986})}\BibitemShut {NoStop}%
\bibitem [{\citenamefont {Barr}\ and\ \citenamefont
  {Seckel}(1992)}]{Barr:1992qq}%
  \BibitemOpen
  \bibfield  {author} {\bibinfo {author} {\bibfnamefont {S.~M.}\ \bibnamefont
  {Barr}}\ and\ \bibinfo {author} {\bibfnamefont {D.}~\bibnamefont {Seckel}},\
  }\href {\doibase 10.1103/PhysRevD.46.539} {\bibfield  {journal} {\bibinfo
  {journal} {Phys. Rev. D}\ }\textbf {\bibinfo {volume} {46}},\ \bibinfo
  {pages} {539} (\bibinfo {year} {1992})}\BibitemShut {NoStop}%
\bibitem [{\citenamefont {Kamionkowski}\ and\ \citenamefont
  {March-Russell}(1992{\natexlab{a}})}]{Kamionkowski:1992mf}%
  \BibitemOpen
  \bibfield  {author} {\bibinfo {author} {\bibfnamefont {M.}~\bibnamefont
  {Kamionkowski}}\ and\ \bibinfo {author} {\bibfnamefont {J.}~\bibnamefont
  {March-Russell}},\ }\href {\doibase 10.1016/0370-2693(92)90492-M} {\bibfield
  {journal} {\bibinfo  {journal} {Phys. Lett. B}\ }\textbf {\bibinfo {volume}
  {282}},\ \bibinfo {pages} {137} (\bibinfo {year} {1992}{\natexlab{a}})},\
  \Eprint {http://arxiv.org/abs/hep-th/9202003} {arXiv:hep-th/9202003}
  \BibitemShut {NoStop}%
\bibitem [{\citenamefont {Kamionkowski}\ and\ \citenamefont
  {March-Russell}(1992{\natexlab{b}})}]{Kamionkowski:1992ax}%
  \BibitemOpen
  \bibfield  {author} {\bibinfo {author} {\bibfnamefont {M.}~\bibnamefont
  {Kamionkowski}}\ and\ \bibinfo {author} {\bibfnamefont {J.}~\bibnamefont
  {March-Russell}},\ }\href {\doibase 10.1103/PhysRevLett.69.1485} {\bibfield
  {journal} {\bibinfo  {journal} {Phys. Rev. Lett.}\ }\textbf {\bibinfo
  {volume} {69}},\ \bibinfo {pages} {1485} (\bibinfo {year}
  {1992}{\natexlab{b}})},\ \Eprint {http://arxiv.org/abs/hep-th/9201063}
  {arXiv:hep-th/9201063} \BibitemShut {NoStop}%
\bibitem [{\citenamefont {Holman}\ \emph {et~al.}(1992)\citenamefont {Holman},
  \citenamefont {Hsu}, \citenamefont {Kephart}, \citenamefont {Kolb},
  \citenamefont {Watkins},\ and\ \citenamefont {Widrow}}]{Holman:1992us}%
  \BibitemOpen
  \bibfield  {author} {\bibinfo {author} {\bibfnamefont {R.}~\bibnamefont
  {Holman}}, \bibinfo {author} {\bibfnamefont {S.~D.~H.}\ \bibnamefont {Hsu}},
  \bibinfo {author} {\bibfnamefont {T.~W.}\ \bibnamefont {Kephart}}, \bibinfo
  {author} {\bibfnamefont {E.~W.}\ \bibnamefont {Kolb}}, \bibinfo {author}
  {\bibfnamefont {R.}~\bibnamefont {Watkins}}, \ and\ \bibinfo {author}
  {\bibfnamefont {L.~M.}\ \bibnamefont {Widrow}},\ }\href {\doibase
  10.1016/0370-2693(92)90491-L} {\bibfield  {journal} {\bibinfo  {journal}
  {Phys. Lett. B}\ }\textbf {\bibinfo {volume} {282}},\ \bibinfo {pages} {132}
  (\bibinfo {year} {1992})},\ \Eprint {http://arxiv.org/abs/hep-ph/9203206}
  {arXiv:hep-ph/9203206} \BibitemShut {NoStop}%
\bibitem [{\citenamefont {Kallosh}\ \emph {et~al.}(1995)\citenamefont
  {Kallosh}, \citenamefont {Linde}, \citenamefont {Linde},\ and\ \citenamefont
  {Susskind}}]{Kallosh:1995hi}%
  \BibitemOpen
  \bibfield  {author} {\bibinfo {author} {\bibfnamefont {R.}~\bibnamefont
  {Kallosh}}, \bibinfo {author} {\bibfnamefont {A.~D.}\ \bibnamefont {Linde}},
  \bibinfo {author} {\bibfnamefont {D.~A.}\ \bibnamefont {Linde}}, \ and\
  \bibinfo {author} {\bibfnamefont {L.}~\bibnamefont {Susskind}},\ }\href
  {\doibase 10.1103/PhysRevD.52.912} {\bibfield  {journal} {\bibinfo  {journal}
  {Phys. Rev. D}\ }\textbf {\bibinfo {volume} {52}},\ \bibinfo {pages} {912}
  (\bibinfo {year} {1995})},\ \Eprint {http://arxiv.org/abs/hep-th/9502069}
  {arXiv:hep-th/9502069} \BibitemShut {NoStop}%
\bibitem [{\citenamefont {Carpenter}\ \emph
  {et~al.}(2009{\natexlab{a}})\citenamefont {Carpenter}, \citenamefont {Dine},\
  and\ \citenamefont {Festuccia}}]{Carpenter:2009zs}%
  \BibitemOpen
  \bibfield  {author} {\bibinfo {author} {\bibfnamefont {L.~M.}\ \bibnamefont
  {Carpenter}}, \bibinfo {author} {\bibfnamefont {M.}~\bibnamefont {Dine}}, \
  and\ \bibinfo {author} {\bibfnamefont {G.}~\bibnamefont {Festuccia}},\ }\href
  {\doibase 10.1103/PhysRevD.80.125017} {\bibfield  {journal} {\bibinfo
  {journal} {Phys. Rev. D}\ }\textbf {\bibinfo {volume} {80}},\ \bibinfo
  {pages} {125017} (\bibinfo {year} {2009}{\natexlab{a}})},\ \Eprint
  {http://arxiv.org/abs/0906.1273} {arXiv:0906.1273 [hep-th]} \BibitemShut
  {NoStop}%
\bibitem [{\citenamefont {Carpenter}\ \emph
  {et~al.}(2009{\natexlab{b}})\citenamefont {Carpenter}, \citenamefont {Dine},
  \citenamefont {Festuccia},\ and\ \citenamefont {Ubaldi}}]{Carpenter:2009sw}%
  \BibitemOpen
  \bibfield  {author} {\bibinfo {author} {\bibfnamefont {L.~M.}\ \bibnamefont
  {Carpenter}}, \bibinfo {author} {\bibfnamefont {M.}~\bibnamefont {Dine}},
  \bibinfo {author} {\bibfnamefont {G.}~\bibnamefont {Festuccia}}, \ and\
  \bibinfo {author} {\bibfnamefont {L.}~\bibnamefont {Ubaldi}},\ }\href
  {\doibase 10.1103/PhysRevD.80.125023} {\bibfield  {journal} {\bibinfo
  {journal} {Phys. Rev. D}\ }\textbf {\bibinfo {volume} {80}},\ \bibinfo
  {pages} {125023} (\bibinfo {year} {2009}{\natexlab{b}})},\ \Eprint
  {http://arxiv.org/abs/0906.5015} {arXiv:0906.5015 [hep-th]} \BibitemShut
  {NoStop}%
\bibitem [{\citenamefont {Banks}\ and\ \citenamefont
  {Seiberg}(2011)}]{Banks:2010zn}%
  \BibitemOpen
  \bibfield  {author} {\bibinfo {author} {\bibfnamefont {T.}~\bibnamefont
  {Banks}}\ and\ \bibinfo {author} {\bibfnamefont {N.}~\bibnamefont
  {Seiberg}},\ }\href {\doibase 10.1103/PhysRevD.83.084019} {\bibfield
  {journal} {\bibinfo  {journal} {Phys. Rev. D}\ }\textbf {\bibinfo {volume}
  {83}},\ \bibinfo {pages} {084019} (\bibinfo {year} {2011})},\ \Eprint
  {http://arxiv.org/abs/1011.5120} {arXiv:1011.5120 [hep-th]} \BibitemShut
  {NoStop}%
\bibitem [{\citenamefont {Witten}(2018)}]{Witten:2017hdv}%
  \BibitemOpen
  \bibfield  {author} {\bibinfo {author} {\bibfnamefont {E.}~\bibnamefont
  {Witten}},\ }\href {\doibase 10.1038/nphys4348} {\bibfield  {journal}
  {\bibinfo  {journal} {Nature Phys.}\ }\textbf {\bibinfo {volume} {14}},\
  \bibinfo {pages} {116} (\bibinfo {year} {2018})},\ \Eprint
  {http://arxiv.org/abs/1710.01791} {arXiv:1710.01791 [hep-th]} \BibitemShut
  {NoStop}%
\bibitem [{\citenamefont {Harlow}\ and\ \citenamefont
  {Ooguri}(2019)}]{Harlow:2018jwu}%
  \BibitemOpen
  \bibfield  {author} {\bibinfo {author} {\bibfnamefont {D.}~\bibnamefont
  {Harlow}}\ and\ \bibinfo {author} {\bibfnamefont {H.}~\bibnamefont
  {Ooguri}},\ }\href {\doibase 10.1103/PhysRevLett.122.191601} {\bibfield
  {journal} {\bibinfo  {journal} {Phys. Rev. Lett.}\ }\textbf {\bibinfo
  {volume} {122}},\ \bibinfo {pages} {191601} (\bibinfo {year} {2019})},\
  \Eprint {http://arxiv.org/abs/1810.05337} {arXiv:1810.05337 [hep-th]}
  \BibitemShut {NoStop}%
\bibitem [{\citenamefont {Harlow}\ and\ \citenamefont
  {Ooguri}(2021)}]{Harlow:2018tng}%
  \BibitemOpen
  \bibfield  {author} {\bibinfo {author} {\bibfnamefont {D.}~\bibnamefont
  {Harlow}}\ and\ \bibinfo {author} {\bibfnamefont {H.}~\bibnamefont
  {Ooguri}},\ }\href {\doibase 10.1007/s00220-021-04040-y} {\bibfield
  {journal} {\bibinfo  {journal} {Commun. Math. Phys.}\ }\textbf {\bibinfo
  {volume} {383}},\ \bibinfo {pages} {1669} (\bibinfo {year} {2021})},\ \Eprint
  {http://arxiv.org/abs/1810.05338} {arXiv:1810.05338 [hep-th]} \BibitemShut
  {NoStop}%
\bibitem [{\citenamefont {Kim}(1985)}]{Kim:1984pt}%
  \BibitemOpen
  \bibfield  {author} {\bibinfo {author} {\bibfnamefont {J.~E.}\ \bibnamefont
  {Kim}},\ }\href {\doibase 10.1103/PhysRevD.31.1733} {\bibfield  {journal}
  {\bibinfo  {journal} {Phys. Rev. D}\ }\textbf {\bibinfo {volume} {31}},\
  \bibinfo {pages} {1733} (\bibinfo {year} {1985})}\BibitemShut {NoStop}%
\bibitem [{\citenamefont {Choi}\ and\ \citenamefont {Kim}(1985)}]{Choi:1985cb}%
  \BibitemOpen
  \bibfield  {author} {\bibinfo {author} {\bibfnamefont {K.}~\bibnamefont
  {Choi}}\ and\ \bibinfo {author} {\bibfnamefont {J.~E.}\ \bibnamefont {Kim}},\
  }\href {\doibase 10.1103/PhysRevD.32.1828} {\bibfield  {journal} {\bibinfo
  {journal} {Phys. Rev. D}\ }\textbf {\bibinfo {volume} {32}},\ \bibinfo
  {pages} {1828} (\bibinfo {year} {1985})}\BibitemShut {NoStop}%
\bibitem [{\citenamefont {Randall}(1992)}]{Randall:1992ut}%
  \BibitemOpen
  \bibfield  {author} {\bibinfo {author} {\bibfnamefont {L.}~\bibnamefont
  {Randall}},\ }\href {\doibase 10.1016/0370-2693(92)91928-3} {\bibfield
  {journal} {\bibinfo  {journal} {Phys. Lett. B}\ }\textbf {\bibinfo {volume}
  {284}},\ \bibinfo {pages} {77} (\bibinfo {year} {1992})}\BibitemShut
  {NoStop}%
\bibitem [{\citenamefont {Izawa}\ \emph {et~al.}(2002)\citenamefont {Izawa},
  \citenamefont {Watari},\ and\ \citenamefont {Yanagida}}]{Izawa:2002qk}%
  \BibitemOpen
  \bibfield  {author} {\bibinfo {author} {\bibfnamefont {K.~I.}\ \bibnamefont
  {Izawa}}, \bibinfo {author} {\bibfnamefont {T.}~\bibnamefont {Watari}}, \
  and\ \bibinfo {author} {\bibfnamefont {T.}~\bibnamefont {Yanagida}},\ }\href
  {\doibase 10.1016/S0370-2693(02)01663-5} {\bibfield  {journal} {\bibinfo
  {journal} {Phys. Lett. B}\ }\textbf {\bibinfo {volume} {534}},\ \bibinfo
  {pages} {93} (\bibinfo {year} {2002})},\ \Eprint
  {http://arxiv.org/abs/hep-ph/0202171} {arXiv:hep-ph/0202171} \BibitemShut
  {NoStop}%
\bibitem [{\citenamefont {Yamada}\ \emph {et~al.}(2016)\citenamefont {Yamada},
  \citenamefont {Yanagida},\ and\ \citenamefont {Yonekura}}]{Yamada:2015waa}%
  \BibitemOpen
  \bibfield  {author} {\bibinfo {author} {\bibfnamefont {M.}~\bibnamefont
  {Yamada}}, \bibinfo {author} {\bibfnamefont {T.~T.}\ \bibnamefont
  {Yanagida}}, \ and\ \bibinfo {author} {\bibfnamefont {K.}~\bibnamefont
  {Yonekura}},\ }\href {\doibase 10.1103/PhysRevLett.116.051801} {\bibfield
  {journal} {\bibinfo  {journal} {Phys. Rev. Lett.}\ }\textbf {\bibinfo
  {volume} {116}},\ \bibinfo {pages} {051801} (\bibinfo {year} {2016})},\
  \Eprint {http://arxiv.org/abs/1510.06504} {arXiv:1510.06504 [hep-ph]}
  \BibitemShut {NoStop}%
\bibitem [{\citenamefont {Redi}\ and\ \citenamefont
  {Sato}(2016)}]{Redi:2016esr}%
  \BibitemOpen
  \bibfield  {author} {\bibinfo {author} {\bibfnamefont {M.}~\bibnamefont
  {Redi}}\ and\ \bibinfo {author} {\bibfnamefont {R.}~\bibnamefont {Sato}},\
  }\href {\doibase 10.1007/JHEP05(2016)104} {\bibfield  {journal} {\bibinfo
  {journal} {JHEP}\ }\textbf {\bibinfo {volume} {05}},\ \bibinfo {pages} {104}
  (\bibinfo {year} {2016})},\ \Eprint {http://arxiv.org/abs/1602.05427}
  {arXiv:1602.05427 [hep-ph]} \BibitemShut {NoStop}%
\bibitem [{\citenamefont {Di~Luzio}\ \emph {et~al.}(2017)\citenamefont
  {Di~Luzio}, \citenamefont {Nardi},\ and\ \citenamefont
  {Ubaldi}}]{DiLuzio:2017tjx}%
  \BibitemOpen
  \bibfield  {author} {\bibinfo {author} {\bibfnamefont {L.}~\bibnamefont
  {Di~Luzio}}, \bibinfo {author} {\bibfnamefont {E.}~\bibnamefont {Nardi}}, \
  and\ \bibinfo {author} {\bibfnamefont {L.}~\bibnamefont {Ubaldi}},\ }\href
  {\doibase 10.1103/PhysRevLett.119.011801} {\bibfield  {journal} {\bibinfo
  {journal} {Phys. Rev. Lett.}\ }\textbf {\bibinfo {volume} {119}},\ \bibinfo
  {pages} {011801} (\bibinfo {year} {2017})},\ \Eprint
  {http://arxiv.org/abs/1704.01122} {arXiv:1704.01122 [hep-ph]} \BibitemShut
  {NoStop}%
\bibitem [{\citenamefont {Lillard}\ and\ \citenamefont
  {Tait}(2017)}]{Lillard:2017cwx}%
  \BibitemOpen
  \bibfield  {author} {\bibinfo {author} {\bibfnamefont {B.}~\bibnamefont
  {Lillard}}\ and\ \bibinfo {author} {\bibfnamefont {T.~M.~P.}\ \bibnamefont
  {Tait}},\ }\href {\doibase 10.1007/JHEP11(2017)005} {\bibfield  {journal}
  {\bibinfo  {journal} {JHEP}\ }\textbf {\bibinfo {volume} {11}},\ \bibinfo
  {pages} {005} (\bibinfo {year} {2017})},\ \Eprint
  {http://arxiv.org/abs/1707.04261} {arXiv:1707.04261 [hep-ph]} \BibitemShut
  {NoStop}%
\bibitem [{\citenamefont {Lillard}\ and\ \citenamefont
  {Tait}(2018)}]{Lillard:2018fdt}%
  \BibitemOpen
  \bibfield  {author} {\bibinfo {author} {\bibfnamefont {B.}~\bibnamefont
  {Lillard}}\ and\ \bibinfo {author} {\bibfnamefont {T.~M.~P.}\ \bibnamefont
  {Tait}},\ }\href {\doibase 10.1007/JHEP11(2018)199} {\bibfield  {journal}
  {\bibinfo  {journal} {JHEP}\ }\textbf {\bibinfo {volume} {11}},\ \bibinfo
  {pages} {199} (\bibinfo {year} {2018})},\ \Eprint
  {http://arxiv.org/abs/1811.03089} {arXiv:1811.03089 [hep-ph]} \BibitemShut
  {NoStop}%
\bibitem [{\citenamefont {Gavela}\ \emph {et~al.}(2019)\citenamefont {Gavela},
  \citenamefont {Ibe}, \citenamefont {Quilez},\ and\ \citenamefont
  {Yanagida}}]{Gavela:2018paw}%
  \BibitemOpen
  \bibfield  {author} {\bibinfo {author} {\bibfnamefont {M.~B.}\ \bibnamefont
  {Gavela}}, \bibinfo {author} {\bibfnamefont {M.}~\bibnamefont {Ibe}},
  \bibinfo {author} {\bibfnamefont {P.}~\bibnamefont {Quilez}}, \ and\ \bibinfo
  {author} {\bibfnamefont {T.~T.}\ \bibnamefont {Yanagida}},\ }\href {\doibase
  10.1140/epjc/s10052-019-7046-3} {\bibfield  {journal} {\bibinfo  {journal}
  {Eur. Phys. J. C}\ }\textbf {\bibinfo {volume} {79}},\ \bibinfo {pages} {542}
  (\bibinfo {year} {2019})},\ \Eprint {http://arxiv.org/abs/1812.08174}
  {arXiv:1812.08174 [hep-ph]} \BibitemShut {NoStop}%
\bibitem [{\citenamefont {Lee}\ and\ \citenamefont {Yin}(2019)}]{Lee:2018yak}%
  \BibitemOpen
  \bibfield  {author} {\bibinfo {author} {\bibfnamefont {H.-S.}\ \bibnamefont
  {Lee}}\ and\ \bibinfo {author} {\bibfnamefont {W.}~\bibnamefont {Yin}},\
  }\href {\doibase 10.1103/PhysRevD.99.015041} {\bibfield  {journal} {\bibinfo
  {journal} {Phys. Rev. D}\ }\textbf {\bibinfo {volume} {99}},\ \bibinfo
  {pages} {015041} (\bibinfo {year} {2019})},\ \Eprint
  {http://arxiv.org/abs/1811.04039} {arXiv:1811.04039 [hep-ph]} \BibitemShut
  {NoStop}%
\bibitem [{\citenamefont {Yamada}\ and\ \citenamefont
  {Yanagida}(2021)}]{Yamada:2021uze}%
  \BibitemOpen
  \bibfield  {author} {\bibinfo {author} {\bibfnamefont {M.}~\bibnamefont
  {Yamada}}\ and\ \bibinfo {author} {\bibfnamefont {T.~T.}\ \bibnamefont
  {Yanagida}},\ }\href {\doibase 10.1016/j.physletb.2021.136267} {\bibfield
  {journal} {\bibinfo  {journal} {Phys. Lett. B}\ }\textbf {\bibinfo {volume}
  {816}},\ \bibinfo {pages} {136267} (\bibinfo {year} {2021})},\ \Eprint
  {http://arxiv.org/abs/2101.10350} {arXiv:2101.10350 [hep-ph]} \BibitemShut
  {NoStop}%
\bibitem [{\citenamefont {Ishida}\ \emph {et~al.}(2022)\citenamefont {Ishida},
  \citenamefont {Matsuzaki},\ and\ \citenamefont {Peng}}]{Ishida:2021avk}%
  \BibitemOpen
  \bibfield  {author} {\bibinfo {author} {\bibfnamefont {H.}~\bibnamefont
  {Ishida}}, \bibinfo {author} {\bibfnamefont {S.}~\bibnamefont {Matsuzaki}}, \
  and\ \bibinfo {author} {\bibfnamefont {X.-C.}\ \bibnamefont {Peng}},\ }\href
  {\doibase 10.1140/epjc/s10052-022-10018-4} {\bibfield  {journal} {\bibinfo
  {journal} {Eur. Phys. J. C}\ }\textbf {\bibinfo {volume} {82}},\ \bibinfo
  {pages} {107} (\bibinfo {year} {2022})},\ \Eprint
  {http://arxiv.org/abs/2103.13644} {arXiv:2103.13644 [hep-ph]} \BibitemShut
  {NoStop}%
\bibitem [{\citenamefont {Contino}\ \emph {et~al.}(2022)\citenamefont
  {Contino}, \citenamefont {Podo},\ and\ \citenamefont
  {Revello}}]{Contino:2021ayn}%
  \BibitemOpen
  \bibfield  {author} {\bibinfo {author} {\bibfnamefont {R.}~\bibnamefont
  {Contino}}, \bibinfo {author} {\bibfnamefont {A.}~\bibnamefont {Podo}}, \
  and\ \bibinfo {author} {\bibfnamefont {F.}~\bibnamefont {Revello}},\ }\href
  {\doibase 10.1007/JHEP04(2022)180} {\bibfield  {journal} {\bibinfo  {journal}
  {JHEP}\ }\textbf {\bibinfo {volume} {04}},\ \bibinfo {pages} {180} (\bibinfo
  {year} {2022})},\ \Eprint {http://arxiv.org/abs/2112.09635} {arXiv:2112.09635
  [hep-ph]} \BibitemShut {NoStop}%
\bibitem [{\citenamefont {Nakai}\ and\ \citenamefont
  {Suzuki}(2021)}]{Nakai:2021nyf}%
  \BibitemOpen
  \bibfield  {author} {\bibinfo {author} {\bibfnamefont {Y.}~\bibnamefont
  {Nakai}}\ and\ \bibinfo {author} {\bibfnamefont {M.}~\bibnamefont {Suzuki}},\
  }\href {\doibase 10.1016/j.physletb.2021.136239} {\bibfield  {journal}
  {\bibinfo  {journal} {Phys. Lett. B}\ }\textbf {\bibinfo {volume} {816}},\
  \bibinfo {pages} {136239} (\bibinfo {year} {2021})},\ \Eprint
  {http://arxiv.org/abs/2102.01329} {arXiv:2102.01329 [hep-ph]} \BibitemShut
  {NoStop}%
\bibitem [{\citenamefont {Nakagawa}\ \emph {et~al.}(2024)\citenamefont
  {Nakagawa}, \citenamefont {Nakai}, \citenamefont {Yamada},\ and\
  \citenamefont {Zhang}}]{Nakagawa:2023shi}%
  \BibitemOpen
  \bibfield  {author} {\bibinfo {author} {\bibfnamefont {S.}~\bibnamefont
  {Nakagawa}}, \bibinfo {author} {\bibfnamefont {Y.}~\bibnamefont {Nakai}},
  \bibinfo {author} {\bibfnamefont {M.}~\bibnamefont {Yamada}}, \ and\ \bibinfo
  {author} {\bibfnamefont {Y.}~\bibnamefont {Zhang}},\ }\href {\doibase
  10.1016/j.physletb.2024.138447} {\bibfield  {journal} {\bibinfo  {journal}
  {Phys. Lett. B}\ }\textbf {\bibinfo {volume} {849}},\ \bibinfo {pages}
  {138447} (\bibinfo {year} {2024})},\ \Eprint
  {http://arxiv.org/abs/2309.06964} {arXiv:2309.06964 [hep-ph]} \BibitemShut
  {NoStop}%
\bibitem [{\citenamefont {Flacke}\ \emph {et~al.}(2007)\citenamefont {Flacke},
  \citenamefont {Gripaios}, \citenamefont {March-Russell},\ and\ \citenamefont
  {Maybury}}]{Flacke:2006ad}%
  \BibitemOpen
  \bibfield  {author} {\bibinfo {author} {\bibfnamefont {T.}~\bibnamefont
  {Flacke}}, \bibinfo {author} {\bibfnamefont {B.}~\bibnamefont {Gripaios}},
  \bibinfo {author} {\bibfnamefont {J.}~\bibnamefont {March-Russell}}, \ and\
  \bibinfo {author} {\bibfnamefont {D.}~\bibnamefont {Maybury}},\ }\href
  {\doibase 10.1088/1126-6708/2007/01/061} {\bibfield  {journal} {\bibinfo
  {journal} {JHEP}\ }\textbf {\bibinfo {volume} {01}},\ \bibinfo {pages} {061}
  (\bibinfo {year} {2007})},\ \Eprint {http://arxiv.org/abs/hep-ph/0611278}
  {arXiv:hep-ph/0611278} \BibitemShut {NoStop}%
\bibitem [{\citenamefont {Cox}\ \emph {et~al.}(2020)\citenamefont {Cox},
  \citenamefont {Gherghetta},\ and\ \citenamefont {Nguyen}}]{Cox:2019rro}%
  \BibitemOpen
  \bibfield  {author} {\bibinfo {author} {\bibfnamefont {P.}~\bibnamefont
  {Cox}}, \bibinfo {author} {\bibfnamefont {T.}~\bibnamefont {Gherghetta}}, \
  and\ \bibinfo {author} {\bibfnamefont {M.~D.}\ \bibnamefont {Nguyen}},\
  }\href {\doibase 10.1007/JHEP01(2020)188} {\bibfield  {journal} {\bibinfo
  {journal} {JHEP}\ }\textbf {\bibinfo {volume} {01}},\ \bibinfo {pages} {188}
  (\bibinfo {year} {2020})},\ \Eprint {http://arxiv.org/abs/1911.09385}
  {arXiv:1911.09385 [hep-ph]} \BibitemShut {NoStop}%
\bibitem [{\citenamefont {Bonnefoy}\ \emph {et~al.}(2021)\citenamefont
  {Bonnefoy}, \citenamefont {Cox}, \citenamefont {Dudas}, \citenamefont
  {Gherghetta},\ and\ \citenamefont {Nguyen}}]{Bonnefoy:2020llz}%
  \BibitemOpen
  \bibfield  {author} {\bibinfo {author} {\bibfnamefont {Q.}~\bibnamefont
  {Bonnefoy}}, \bibinfo {author} {\bibfnamefont {P.}~\bibnamefont {Cox}},
  \bibinfo {author} {\bibfnamefont {E.}~\bibnamefont {Dudas}}, \bibinfo
  {author} {\bibfnamefont {T.}~\bibnamefont {Gherghetta}}, \ and\ \bibinfo
  {author} {\bibfnamefont {M.~D.}\ \bibnamefont {Nguyen}},\ }\href {\doibase
  10.1007/JHEP04(2021)084} {\bibfield  {journal} {\bibinfo  {journal} {JHEP}\
  }\textbf {\bibinfo {volume} {04}},\ \bibinfo {pages} {084} (\bibinfo {year}
  {2021})},\ \Eprint {http://arxiv.org/abs/2012.09728} {arXiv:2012.09728
  [hep-ph]} \BibitemShut {NoStop}%
\bibitem [{\citenamefont {Lee}\ \emph {et~al.}(2022)\citenamefont {Lee},
  \citenamefont {Nakai},\ and\ \citenamefont {Suzuki}}]{Lee:2021slp}%
  \BibitemOpen
  \bibfield  {author} {\bibinfo {author} {\bibfnamefont {S.~J.}\ \bibnamefont
  {Lee}}, \bibinfo {author} {\bibfnamefont {Y.}~\bibnamefont {Nakai}}, \ and\
  \bibinfo {author} {\bibfnamefont {M.}~\bibnamefont {Suzuki}},\ }\href
  {\doibase 10.1007/JHEP03(2022)038} {\bibfield  {journal} {\bibinfo  {journal}
  {JHEP}\ }\textbf {\bibinfo {volume} {03}},\ \bibinfo {pages} {038} (\bibinfo
  {year} {2022})},\ \Eprint {http://arxiv.org/abs/2112.08083} {arXiv:2112.08083
  [hep-ph]} \BibitemShut {NoStop}%
\bibitem [{\citenamefont {Rubakov}(1997)}]{Rubakov:1997vp}%
  \BibitemOpen
  \bibfield  {author} {\bibinfo {author} {\bibfnamefont {V.~A.}\ \bibnamefont
  {Rubakov}},\ }\href {\doibase 10.1134/1.567390} {\bibfield  {journal}
  {\bibinfo  {journal} {JETP Lett.}\ }\textbf {\bibinfo {volume} {65}},\
  \bibinfo {pages} {621} (\bibinfo {year} {1997})},\ \Eprint
  {http://arxiv.org/abs/hep-ph/9703409} {arXiv:hep-ph/9703409} \BibitemShut
  {NoStop}%
\bibitem [{\citenamefont {Berezhiani}\ \emph {et~al.}(2001)\citenamefont
  {Berezhiani}, \citenamefont {Gianfagna},\ and\ \citenamefont
  {Giannotti}}]{Berezhiani:2000gh}%
  \BibitemOpen
  \bibfield  {author} {\bibinfo {author} {\bibfnamefont {Z.}~\bibnamefont
  {Berezhiani}}, \bibinfo {author} {\bibfnamefont {L.}~\bibnamefont
  {Gianfagna}}, \ and\ \bibinfo {author} {\bibfnamefont {M.}~\bibnamefont
  {Giannotti}},\ }\href {\doibase 10.1016/S0370-2693(00)01392-7} {\bibfield
  {journal} {\bibinfo  {journal} {Phys. Lett. B}\ }\textbf {\bibinfo {volume}
  {500}},\ \bibinfo {pages} {286} (\bibinfo {year} {2001})},\ \Eprint
  {http://arxiv.org/abs/hep-ph/0009290} {arXiv:hep-ph/0009290} \BibitemShut
  {NoStop}%
\bibitem [{\citenamefont {Hook}(2015)}]{Hook:2014cda}%
  \BibitemOpen
  \bibfield  {author} {\bibinfo {author} {\bibfnamefont {A.}~\bibnamefont
  {Hook}},\ }\href {\doibase 10.1103/PhysRevLett.114.141801} {\bibfield
  {journal} {\bibinfo  {journal} {Phys. Rev. Lett.}\ }\textbf {\bibinfo
  {volume} {114}},\ \bibinfo {pages} {141801} (\bibinfo {year} {2015})},\
  \Eprint {http://arxiv.org/abs/1411.3325} {arXiv:1411.3325 [hep-ph]}
  \BibitemShut {NoStop}%
\bibitem [{\citenamefont {Fukuda}\ \emph {et~al.}(2015)\citenamefont {Fukuda},
  \citenamefont {Harigaya}, \citenamefont {Ibe},\ and\ \citenamefont
  {Yanagida}}]{Fukuda:2015ana}%
  \BibitemOpen
  \bibfield  {author} {\bibinfo {author} {\bibfnamefont {H.}~\bibnamefont
  {Fukuda}}, \bibinfo {author} {\bibfnamefont {K.}~\bibnamefont {Harigaya}},
  \bibinfo {author} {\bibfnamefont {M.}~\bibnamefont {Ibe}}, \ and\ \bibinfo
  {author} {\bibfnamefont {T.~T.}\ \bibnamefont {Yanagida}},\ }\href {\doibase
  10.1103/PhysRevD.92.015021} {\bibfield  {journal} {\bibinfo  {journal} {Phys.
  Rev. D}\ }\textbf {\bibinfo {volume} {92}},\ \bibinfo {pages} {015021}
  (\bibinfo {year} {2015})},\ \Eprint {http://arxiv.org/abs/1504.06084}
  {arXiv:1504.06084 [hep-ph]} \BibitemShut {NoStop}%
\bibitem [{\citenamefont {Gherghetta}\ \emph {et~al.}(2016)\citenamefont
  {Gherghetta}, \citenamefont {Nagata},\ and\ \citenamefont
  {Shifman}}]{Gherghetta:2016fhp}%
  \BibitemOpen
  \bibfield  {author} {\bibinfo {author} {\bibfnamefont {T.}~\bibnamefont
  {Gherghetta}}, \bibinfo {author} {\bibfnamefont {N.}~\bibnamefont {Nagata}},
  \ and\ \bibinfo {author} {\bibfnamefont {M.}~\bibnamefont {Shifman}},\ }\href
  {\doibase 10.1103/PhysRevD.93.115010} {\bibfield  {journal} {\bibinfo
  {journal} {Phys. Rev. D}\ }\textbf {\bibinfo {volume} {93}},\ \bibinfo
  {pages} {115010} (\bibinfo {year} {2016})},\ \Eprint
  {http://arxiv.org/abs/1604.01127} {arXiv:1604.01127 [hep-ph]} \BibitemShut
  {NoStop}%
\bibitem [{\citenamefont {Dimopoulos}\ \emph {et~al.}(2016)\citenamefont
  {Dimopoulos}, \citenamefont {Hook}, \citenamefont {Huang},\ and\
  \citenamefont {Marques-Tavares}}]{Dimopoulos:2016lvn}%
  \BibitemOpen
  \bibfield  {author} {\bibinfo {author} {\bibfnamefont {S.}~\bibnamefont
  {Dimopoulos}}, \bibinfo {author} {\bibfnamefont {A.}~\bibnamefont {Hook}},
  \bibinfo {author} {\bibfnamefont {J.}~\bibnamefont {Huang}}, \ and\ \bibinfo
  {author} {\bibfnamefont {G.}~\bibnamefont {Marques-Tavares}},\ }\href
  {\doibase 10.1007/JHEP11(2016)052} {\bibfield  {journal} {\bibinfo  {journal}
  {JHEP}\ }\textbf {\bibinfo {volume} {11}},\ \bibinfo {pages} {052} (\bibinfo
  {year} {2016})},\ \Eprint {http://arxiv.org/abs/1606.03097} {arXiv:1606.03097
  [hep-ph]} \BibitemShut {NoStop}%
\bibitem [{\citenamefont {Gherghetta}\ and\ \citenamefont
  {Nguyen}(2020)}]{Gherghetta:2020ofz}%
  \BibitemOpen
  \bibfield  {author} {\bibinfo {author} {\bibfnamefont {T.}~\bibnamefont
  {Gherghetta}}\ and\ \bibinfo {author} {\bibfnamefont {M.~D.}\ \bibnamefont
  {Nguyen}},\ }\href {\doibase 10.1007/JHEP12(2020)094} {\bibfield  {journal}
  {\bibinfo  {journal} {JHEP}\ }\textbf {\bibinfo {volume} {12}},\ \bibinfo
  {pages} {094} (\bibinfo {year} {2020})},\ \Eprint
  {http://arxiv.org/abs/2007.10875} {arXiv:2007.10875 [hep-ph]} \BibitemShut
  {NoStop}%
\bibitem [{\citenamefont {Alves}\ and\ \citenamefont
  {Weiner}(2018)}]{Alves:2017avw}%
  \BibitemOpen
  \bibfield  {author} {\bibinfo {author} {\bibfnamefont {D.~S.~M.}\
  \bibnamefont {Alves}}\ and\ \bibinfo {author} {\bibfnamefont
  {N.}~\bibnamefont {Weiner}},\ }\href {\doibase 10.1007/JHEP07(2018)092}
  {\bibfield  {journal} {\bibinfo  {journal} {JHEP}\ }\textbf {\bibinfo
  {volume} {07}},\ \bibinfo {pages} {092} (\bibinfo {year} {2018})},\ \Eprint
  {http://arxiv.org/abs/1710.03764} {arXiv:1710.03764 [hep-ph]} \BibitemShut
  {NoStop}%
\bibitem [{\citenamefont {Liu}\ \emph {et~al.}(2021)\citenamefont {Liu},
  \citenamefont {McGinnis}, \citenamefont {Wagner},\ and\ \citenamefont
  {Wang}}]{Liu:2021wap}%
  \BibitemOpen
  \bibfield  {author} {\bibinfo {author} {\bibfnamefont {J.}~\bibnamefont
  {Liu}}, \bibinfo {author} {\bibfnamefont {N.}~\bibnamefont {McGinnis}},
  \bibinfo {author} {\bibfnamefont {C.~E.~M.}\ \bibnamefont {Wagner}}, \ and\
  \bibinfo {author} {\bibfnamefont {X.-P.}\ \bibnamefont {Wang}},\ }\href
  {\doibase 10.1007/JHEP05(2021)138} {\bibfield  {journal} {\bibinfo  {journal}
  {JHEP}\ }\textbf {\bibinfo {volume} {05}},\ \bibinfo {pages} {138} (\bibinfo
  {year} {2021})},\ \Eprint {http://arxiv.org/abs/2102.10118} {arXiv:2102.10118
  [hep-ph]} \BibitemShut {NoStop}%
\bibitem [{\citenamefont {Girmohanta}\ \emph {et~al.}(2024)\citenamefont
  {Girmohanta}, \citenamefont {Nakagawa}, \citenamefont {Nakai},\ and\
  \citenamefont {Xu}}]{Girmohanta:2024nyf}%
  \BibitemOpen
  \bibfield  {author} {\bibinfo {author} {\bibfnamefont {S.}~\bibnamefont
  {Girmohanta}}, \bibinfo {author} {\bibfnamefont {S.}~\bibnamefont
  {Nakagawa}}, \bibinfo {author} {\bibfnamefont {Y.}~\bibnamefont {Nakai}}, \
  and\ \bibinfo {author} {\bibfnamefont {J.}~\bibnamefont {Xu}},\ }\href
  {\doibase 10.1007/JHEP10(2024)153} {\bibfield  {journal} {\bibinfo  {journal}
  {JHEP}\ }\textbf {\bibinfo {volume} {10}},\ \bibinfo {pages} {153} (\bibinfo
  {year} {2024})},\ \Eprint {http://arxiv.org/abs/2405.13425} {arXiv:2405.13425
  [hep-ph]} \BibitemShut {NoStop}%
\bibitem [{\citenamefont {Cheng}\ and\ \citenamefont
  {Kaplan}(2001)}]{Cheng:2001ys}%
  \BibitemOpen
  \bibfield  {author} {\bibinfo {author} {\bibfnamefont {H.-C.}\ \bibnamefont
  {Cheng}}\ and\ \bibinfo {author} {\bibfnamefont {D.~E.}\ \bibnamefont
  {Kaplan}},\ }\href@noop {} {\  (\bibinfo {year} {2001})},\ \Eprint
  {http://arxiv.org/abs/hep-ph/0103346} {arXiv:hep-ph/0103346} \BibitemShut
  {NoStop}%
\bibitem [{\citenamefont {Harigaya}\ \emph {et~al.}(2013)\citenamefont
  {Harigaya}, \citenamefont {Ibe}, \citenamefont {Schmitz},\ and\ \citenamefont
  {Yanagida}}]{Harigaya:2013vja}%
  \BibitemOpen
  \bibfield  {author} {\bibinfo {author} {\bibfnamefont {K.}~\bibnamefont
  {Harigaya}}, \bibinfo {author} {\bibfnamefont {M.}~\bibnamefont {Ibe}},
  \bibinfo {author} {\bibfnamefont {K.}~\bibnamefont {Schmitz}}, \ and\
  \bibinfo {author} {\bibfnamefont {T.~T.}\ \bibnamefont {Yanagida}},\ }\href
  {\doibase 10.1103/PhysRevD.88.075022} {\bibfield  {journal} {\bibinfo
  {journal} {Phys. Rev. D}\ }\textbf {\bibinfo {volume} {88}},\ \bibinfo
  {pages} {075022} (\bibinfo {year} {2013})},\ \Eprint
  {http://arxiv.org/abs/1308.1227} {arXiv:1308.1227 [hep-ph]} \BibitemShut
  {NoStop}%
\bibitem [{\citenamefont {Fukuda}\ \emph {et~al.}(2017)\citenamefont {Fukuda},
  \citenamefont {Ibe}, \citenamefont {Suzuki},\ and\ \citenamefont
  {Yanagida}}]{Fukuda:2017ylt}%
  \BibitemOpen
  \bibfield  {author} {\bibinfo {author} {\bibfnamefont {H.}~\bibnamefont
  {Fukuda}}, \bibinfo {author} {\bibfnamefont {M.}~\bibnamefont {Ibe}},
  \bibinfo {author} {\bibfnamefont {M.}~\bibnamefont {Suzuki}}, \ and\ \bibinfo
  {author} {\bibfnamefont {T.~T.}\ \bibnamefont {Yanagida}},\ }\href {\doibase
  10.1016/j.physletb.2017.05.071} {\bibfield  {journal} {\bibinfo  {journal}
  {Phys. Lett. B}\ }\textbf {\bibinfo {volume} {771}},\ \bibinfo {pages} {327}
  (\bibinfo {year} {2017})},\ \Eprint {http://arxiv.org/abs/1703.01112}
  {arXiv:1703.01112 [hep-ph]} \BibitemShut {NoStop}%
\bibitem [{\citenamefont {Fukuda}\ \emph {et~al.}(2018)\citenamefont {Fukuda},
  \citenamefont {Ibe}, \citenamefont {Suzuki},\ and\ \citenamefont
  {Yanagida}}]{Fukuda:2018oco}%
  \BibitemOpen
  \bibfield  {author} {\bibinfo {author} {\bibfnamefont {H.}~\bibnamefont
  {Fukuda}}, \bibinfo {author} {\bibfnamefont {M.}~\bibnamefont {Ibe}},
  \bibinfo {author} {\bibfnamefont {M.}~\bibnamefont {Suzuki}}, \ and\ \bibinfo
  {author} {\bibfnamefont {T.~T.}\ \bibnamefont {Yanagida}},\ }\href {\doibase
  10.1007/JHEP07(2018)128} {\bibfield  {journal} {\bibinfo  {journal} {JHEP}\
  }\textbf {\bibinfo {volume} {07}},\ \bibinfo {pages} {128} (\bibinfo {year}
  {2018})},\ \Eprint {http://arxiv.org/abs/1803.00759} {arXiv:1803.00759
  [hep-ph]} \BibitemShut {NoStop}%
\bibitem [{\citenamefont {Ibe}\ \emph {et~al.}(2018)\citenamefont {Ibe},
  \citenamefont {Suzuki},\ and\ \citenamefont {Yanagida}}]{Ibe:2018hir}%
  \BibitemOpen
  \bibfield  {author} {\bibinfo {author} {\bibfnamefont {M.}~\bibnamefont
  {Ibe}}, \bibinfo {author} {\bibfnamefont {M.}~\bibnamefont {Suzuki}}, \ and\
  \bibinfo {author} {\bibfnamefont {T.~T.}\ \bibnamefont {Yanagida}},\ }\href
  {\doibase 10.1007/JHEP08(2018)049} {\bibfield  {journal} {\bibinfo  {journal}
  {JHEP}\ }\textbf {\bibinfo {volume} {08}},\ \bibinfo {pages} {049} (\bibinfo
  {year} {2018})},\ \Eprint {http://arxiv.org/abs/1805.10029} {arXiv:1805.10029
  [hep-ph]} \BibitemShut {NoStop}%
\bibitem [{\citenamefont {Choi}\ \emph {et~al.}(2020)\citenamefont {Choi},
  \citenamefont {Suzuki},\ and\ \citenamefont {Yanagida}}]{Choi:2020vgb}%
  \BibitemOpen
  \bibfield  {author} {\bibinfo {author} {\bibfnamefont {G.}~\bibnamefont
  {Choi}}, \bibinfo {author} {\bibfnamefont {M.}~\bibnamefont {Suzuki}}, \ and\
  \bibinfo {author} {\bibfnamefont {T.~T.}\ \bibnamefont {Yanagida}},\ }\href
  {\doibase 10.1007/JHEP07(2020)048} {\bibfield  {journal} {\bibinfo  {journal}
  {JHEP}\ }\textbf {\bibinfo {volume} {07}},\ \bibinfo {pages} {048} (\bibinfo
  {year} {2020})},\ \Eprint {http://arxiv.org/abs/2005.10415} {arXiv:2005.10415
  [hep-ph]} \BibitemShut {NoStop}%
\bibitem [{\citenamefont {Yin}(2020)}]{Yin:2020dfn}%
  \BibitemOpen
  \bibfield  {author} {\bibinfo {author} {\bibfnamefont {W.}~\bibnamefont
  {Yin}},\ }\href {\doibase 10.1007/JHEP10(2020)032} {\bibfield  {journal}
  {\bibinfo  {journal} {JHEP}\ }\textbf {\bibinfo {volume} {10}},\ \bibinfo
  {pages} {032} (\bibinfo {year} {2020})},\ \Eprint
  {http://arxiv.org/abs/2007.13320} {arXiv:2007.13320 [hep-ph]} \BibitemShut
  {NoStop}%
\bibitem [{\citenamefont {Chen}\ \emph {et~al.}(2021)\citenamefont {Chen},
  \citenamefont {Liu},\ and\ \citenamefont {Teng}}]{Chen:2021haa}%
  \BibitemOpen
  \bibfield  {author} {\bibinfo {author} {\bibfnamefont {N.}~\bibnamefont
  {Chen}}, \bibinfo {author} {\bibfnamefont {Y.}~\bibnamefont {Liu}}, \ and\
  \bibinfo {author} {\bibfnamefont {Z.}~\bibnamefont {Teng}},\ }\href {\doibase
  10.1103/PhysRevD.104.115011} {\bibfield  {journal} {\bibinfo  {journal}
  {Phys. Rev. D}\ }\textbf {\bibinfo {volume} {104}},\ \bibinfo {pages}
  {115011} (\bibinfo {year} {2021})},\ \Eprint
  {http://arxiv.org/abs/2106.00223} {arXiv:2106.00223 [hep-ph]} \BibitemShut
  {NoStop}%
\bibitem [{\citenamefont {Seiberg}(1995)}]{Seiberg:1994pq}%
  \BibitemOpen
  \bibfield  {author} {\bibinfo {author} {\bibfnamefont {N.}~\bibnamefont
  {Seiberg}},\ }\href {\doibase 10.1016/0550-3213(94)00023-8} {\bibfield
  {journal} {\bibinfo  {journal} {Nucl. Phys. B}\ }\textbf {\bibinfo {volume}
  {435}},\ \bibinfo {pages} {129} (\bibinfo {year} {1995})},\ \Eprint
  {http://arxiv.org/abs/hep-th/9411149} {arXiv:hep-th/9411149} \BibitemShut
  {NoStop}%
\bibitem [{\citenamefont {Cs\'aki}\ \emph {et~al.}(2024)\citenamefont
  {Cs\'aki}, \citenamefont {Ovadia}, \citenamefont {Ruhdorfer}, \citenamefont
  {Telem},\ and\ \citenamefont {Terning}}]{Csaki:2024plt}%
  \BibitemOpen
  \bibfield  {author} {\bibinfo {author} {\bibfnamefont {C.}~\bibnamefont
  {Cs\'aki}}, \bibinfo {author} {\bibfnamefont {R.}~\bibnamefont {Ovadia}},
  \bibinfo {author} {\bibfnamefont {M.}~\bibnamefont {Ruhdorfer}}, \bibinfo
  {author} {\bibfnamefont {O.}~\bibnamefont {Telem}}, \ and\ \bibinfo {author}
  {\bibfnamefont {J.}~\bibnamefont {Terning}},\ }\href@noop {} {\  (\bibinfo
  {year} {2024})},\ \Eprint {http://arxiv.org/abs/2411.15312} {arXiv:2411.15312
  [hep-ph]} \BibitemShut {NoStop}%
\bibitem [{\citenamefont {Seiberg}\ and\ \citenamefont
  {Witten}(1994)}]{Seiberg:1994rs}%
  \BibitemOpen
  \bibfield  {author} {\bibinfo {author} {\bibfnamefont {N.}~\bibnamefont
  {Seiberg}}\ and\ \bibinfo {author} {\bibfnamefont {E.}~\bibnamefont
  {Witten}},\ }\href {\doibase 10.1016/0550-3213(94)90124-4} {\bibfield
  {journal} {\bibinfo  {journal} {Nucl. Phys. B}\ }\textbf {\bibinfo {volume}
  {426}},\ \bibinfo {pages} {19} (\bibinfo {year} {1994})},\ \bibinfo {note}
  {[Erratum: Nucl.Phys.B 430, 485--486 (1994)]},\ \Eprint
  {http://arxiv.org/abs/hep-th/9407087} {arXiv:hep-th/9407087} \BibitemShut
  {NoStop}%
\bibitem [{\citenamefont {Intriligator}\ and\ \citenamefont
  {Seiberg}(2007)}]{Intriligator:2007cp}%
  \BibitemOpen
  \bibfield  {author} {\bibinfo {author} {\bibfnamefont {K.~A.}\ \bibnamefont
  {Intriligator}}\ and\ \bibinfo {author} {\bibfnamefont {N.}~\bibnamefont
  {Seiberg}},\ }\href {\doibase 10.1088/0264-9381/24/21/S02} {\bibfield
  {journal} {\bibinfo  {journal} {Class. Quant. Grav.}\ }\textbf {\bibinfo
  {volume} {24}},\ \bibinfo {pages} {S741} (\bibinfo {year} {2007})},\ \Eprint
  {http://arxiv.org/abs/hep-ph/0702069} {arXiv:hep-ph/0702069} \BibitemShut
  {NoStop}%
\bibitem [{\citenamefont {'t~Hooft}(1980)}]{tHooft:1979rat}%
  \BibitemOpen
  \bibfield  {author} {\bibinfo {author} {\bibfnamefont {G.}~\bibnamefont
  {'t~Hooft}},\ }\href {\doibase 10.1007/978-1-4684-7571-5_9} {\bibfield
  {journal} {\bibinfo  {journal} {NATO Sci. Ser. B}\ }\textbf {\bibinfo
  {volume} {59}},\ \bibinfo {pages} {135} (\bibinfo {year} {1980})}\BibitemShut
  {NoStop}%
\bibitem [{\citenamefont {Kim}(1979)}]{Kim:1979if}%
  \BibitemOpen
  \bibfield  {author} {\bibinfo {author} {\bibfnamefont {J.~E.}\ \bibnamefont
  {Kim}},\ }\href {\doibase 10.1103/PhysRevLett.43.103} {\bibfield  {journal}
  {\bibinfo  {journal} {Phys. Rev. Lett.}\ }\textbf {\bibinfo {volume} {43}},\
  \bibinfo {pages} {103} (\bibinfo {year} {1979})}\BibitemShut {NoStop}%
\bibitem [{\citenamefont {Shifman}\ \emph {et~al.}(1980)\citenamefont
  {Shifman}, \citenamefont {Vainshtein},\ and\ \citenamefont
  {Zakharov}}]{Shifman:1979if}%
  \BibitemOpen
  \bibfield  {author} {\bibinfo {author} {\bibfnamefont {M.~A.}\ \bibnamefont
  {Shifman}}, \bibinfo {author} {\bibfnamefont {A.~I.}\ \bibnamefont
  {Vainshtein}}, \ and\ \bibinfo {author} {\bibfnamefont {V.~I.}\ \bibnamefont
  {Zakharov}},\ }\href {\doibase 10.1016/0550-3213(80)90209-6} {\bibfield
  {journal} {\bibinfo  {journal} {Nucl. Phys. B}\ }\textbf {\bibinfo {volume}
  {166}},\ \bibinfo {pages} {493} (\bibinfo {year} {1980})}\BibitemShut
  {NoStop}%
\bibitem [{\citenamefont {Workman}\ \emph {et~al.}(2022)\citenamefont {Workman}
  \emph {et~al.}}]{ParticleDataGroup:2022pth}%
  \BibitemOpen
  \bibfield  {author} {\bibinfo {author} {\bibfnamefont {R.~L.}\ \bibnamefont
  {Workman}} \emph {et~al.} (\bibinfo {collaboration} {Particle Data Group}),\
  }\href {\doibase 10.1093/ptep/ptac097} {\bibfield  {journal} {\bibinfo
  {journal} {PTEP}\ }\textbf {\bibinfo {volume} {2022}},\ \bibinfo {pages}
  {083C01} (\bibinfo {year} {2022})}\BibitemShut {NoStop}%
\bibitem [{\citenamefont {Staub}(2008)}]{Staub:2008uz}%
  \BibitemOpen
  \bibfield  {author} {\bibinfo {author} {\bibfnamefont {F.}~\bibnamefont
  {Staub}},\ }\href@noop {} {\  (\bibinfo {year} {2008})},\ \Eprint
  {http://arxiv.org/abs/0806.0538} {arXiv:0806.0538 [hep-ph]} \BibitemShut
  {NoStop}%
\bibitem [{\citenamefont {Coleman}\ and\ \citenamefont
  {Weinberg}(1973)}]{Coleman:1973jx}%
  \BibitemOpen
  \bibfield  {author} {\bibinfo {author} {\bibfnamefont {S.~R.}\ \bibnamefont
  {Coleman}}\ and\ \bibinfo {author} {\bibfnamefont {E.~J.}\ \bibnamefont
  {Weinberg}},\ }\href {\doibase 10.1103/PhysRevD.7.1888} {\bibfield  {journal}
  {\bibinfo  {journal} {Phys. Rev. D}\ }\textbf {\bibinfo {volume} {7}},\
  \bibinfo {pages} {1888} (\bibinfo {year} {1973})}\BibitemShut {NoStop}%
\bibitem [{\citenamefont {Grilli~di Cortona}\ \emph {et~al.}(2016)\citenamefont
  {Grilli~di Cortona}, \citenamefont {Hardy}, \citenamefont {Pardo~Vega},\ and\
  \citenamefont {Villadoro}}]{GrillidiCortona:2015jxo}%
  \BibitemOpen
  \bibfield  {author} {\bibinfo {author} {\bibfnamefont {G.}~\bibnamefont
  {Grilli~di Cortona}}, \bibinfo {author} {\bibfnamefont {E.}~\bibnamefont
  {Hardy}}, \bibinfo {author} {\bibfnamefont {J.}~\bibnamefont {Pardo~Vega}}, \
  and\ \bibinfo {author} {\bibfnamefont {G.}~\bibnamefont {Villadoro}},\ }\href
  {\doibase 10.1007/JHEP01(2016)034} {\bibfield  {journal} {\bibinfo  {journal}
  {JHEP}\ }\textbf {\bibinfo {volume} {01}},\ \bibinfo {pages} {034} (\bibinfo
  {year} {2016})},\ \Eprint {http://arxiv.org/abs/1511.02867} {arXiv:1511.02867
  [hep-ph]} \BibitemShut {NoStop}%
\bibitem [{\citenamefont {Abel}\ \emph {et~al.}(2020)\citenamefont {Abel} \emph
  {et~al.}}]{Abel:2020pzs}%
  \BibitemOpen
  \bibfield  {author} {\bibinfo {author} {\bibfnamefont {C.}~\bibnamefont
  {Abel}} \emph {et~al.},\ }\href {\doibase 10.1103/PhysRevLett.124.081803}
  {\bibfield  {journal} {\bibinfo  {journal} {Phys. Rev. Lett.}\ }\textbf
  {\bibinfo {volume} {124}},\ \bibinfo {pages} {081803} (\bibinfo {year}
  {2020})},\ \Eprint {http://arxiv.org/abs/2001.11966} {arXiv:2001.11966
  [hep-ex]} \BibitemShut {NoStop}%
\bibitem [{\citenamefont {Ballesteros}\ \emph {et~al.}(2017)\citenamefont
  {Ballesteros}, \citenamefont {Redondo}, \citenamefont {Ringwald},\ and\
  \citenamefont {Tamarit}}]{Ballesteros:2016xej}%
  \BibitemOpen
  \bibfield  {author} {\bibinfo {author} {\bibfnamefont {G.}~\bibnamefont
  {Ballesteros}}, \bibinfo {author} {\bibfnamefont {J.}~\bibnamefont
  {Redondo}}, \bibinfo {author} {\bibfnamefont {A.}~\bibnamefont {Ringwald}}, \
  and\ \bibinfo {author} {\bibfnamefont {C.}~\bibnamefont {Tamarit}},\ }\href
  {\doibase 10.1088/1475-7516/2017/08/001} {\bibfield  {journal} {\bibinfo
  {journal} {JCAP}\ }\textbf {\bibinfo {volume} {08}},\ \bibinfo {pages} {001}
  (\bibinfo {year} {2017})},\ \Eprint {http://arxiv.org/abs/1610.01639}
  {arXiv:1610.01639 [hep-ph]} \BibitemShut {NoStop}%
\bibitem [{\citenamefont {Ahmed}\ \emph
  {et~al.}(2019{\natexlab{a}})\citenamefont {Ahmed} \emph
  {et~al.}}]{TUCAN:2018vmr}%
  \BibitemOpen
  \bibfield  {author} {\bibinfo {author} {\bibfnamefont {S.}~\bibnamefont
  {Ahmed}} \emph {et~al.} (\bibinfo {collaboration} {TUCAN}),\ }\href {\doibase
  10.1103/PhysRevC.99.025503} {\bibfield  {journal} {\bibinfo  {journal} {Phys.
  Rev. C}\ }\textbf {\bibinfo {volume} {99}},\ \bibinfo {pages} {025503}
  (\bibinfo {year} {2019}{\natexlab{a}})},\ \Eprint
  {http://arxiv.org/abs/1809.04071} {arXiv:1809.04071 [physics.ins-det]}
  \BibitemShut {NoStop}%
\bibitem [{\citenamefont {Ahmed}\ \emph
  {et~al.}(2019{\natexlab{b}})\citenamefont {Ahmed} \emph
  {et~al.}}]{nEDM:2019qgk}%
  \BibitemOpen
  \bibfield  {author} {\bibinfo {author} {\bibfnamefont {M.~W.}\ \bibnamefont
  {Ahmed}} \emph {et~al.} (\bibinfo {collaboration} {nEDM}),\ }\href {\doibase
  10.1088/1748-0221/14/11/P11017} {\bibfield  {journal} {\bibinfo  {journal}
  {JINST}\ }\textbf {\bibinfo {volume} {14}},\ \bibinfo {pages} {P11017}
  (\bibinfo {year} {2019}{\natexlab{b}})},\ \Eprint
  {http://arxiv.org/abs/1908.09937} {arXiv:1908.09937 [physics.ins-det]}
  \BibitemShut {NoStop}%
\bibitem [{\citenamefont {Ayres}\ \emph {et~al.}(2021)\citenamefont {Ayres}
  \emph {et~al.}}]{n2EDM:2021yah}%
  \BibitemOpen
  \bibfield  {author} {\bibinfo {author} {\bibfnamefont {N.~J.}\ \bibnamefont
  {Ayres}} \emph {et~al.} (\bibinfo {collaboration} {n2EDM}),\ }\href {\doibase
  10.1140/epjc/s10052-021-09298-z} {\bibfield  {journal} {\bibinfo  {journal}
  {Eur. Phys. J. C}\ }\textbf {\bibinfo {volume} {81}},\ \bibinfo {pages} {512}
  (\bibinfo {year} {2021})},\ \Eprint {http://arxiv.org/abs/2101.08730}
  {arXiv:2101.08730 [physics.ins-det]} \BibitemShut {NoStop}%
\bibitem [{\citenamefont {Jungman}\ \emph {et~al.}(1996)\citenamefont
  {Jungman}, \citenamefont {Kamionkowski},\ and\ \citenamefont
  {Griest}}]{Jungman:1995df}%
  \BibitemOpen
  \bibfield  {author} {\bibinfo {author} {\bibfnamefont {G.}~\bibnamefont
  {Jungman}}, \bibinfo {author} {\bibfnamefont {M.}~\bibnamefont
  {Kamionkowski}}, \ and\ \bibinfo {author} {\bibfnamefont {K.}~\bibnamefont
  {Griest}},\ }\href {\doibase 10.1016/0370-1573(95)00058-5} {\bibfield
  {journal} {\bibinfo  {journal} {Phys. Rept.}\ }\textbf {\bibinfo {volume}
  {267}},\ \bibinfo {pages} {195} (\bibinfo {year} {1996})},\ \Eprint
  {http://arxiv.org/abs/hep-ph/9506380} {arXiv:hep-ph/9506380} \BibitemShut
  {NoStop}%
\bibitem [{\citenamefont {Ibe}\ and\ \citenamefont
  {Yanagida}(2012)}]{Ibe:2011aa}%
  \BibitemOpen
  \bibfield  {author} {\bibinfo {author} {\bibfnamefont {M.}~\bibnamefont
  {Ibe}}\ and\ \bibinfo {author} {\bibfnamefont {T.~T.}\ \bibnamefont
  {Yanagida}},\ }\href {\doibase 10.1016/j.physletb.2012.02.034} {\bibfield
  {journal} {\bibinfo  {journal} {Phys. Lett. B}\ }\textbf {\bibinfo {volume}
  {709}},\ \bibinfo {pages} {374} (\bibinfo {year} {2012})},\ \Eprint
  {http://arxiv.org/abs/1112.2462} {arXiv:1112.2462 [hep-ph]} \BibitemShut
  {NoStop}%
\bibitem [{\citenamefont {Ibe}\ \emph {et~al.}(2012)\citenamefont {Ibe},
  \citenamefont {Matsumoto},\ and\ \citenamefont {Yanagida}}]{Ibe:2012hu}%
  \BibitemOpen
  \bibfield  {author} {\bibinfo {author} {\bibfnamefont {M.}~\bibnamefont
  {Ibe}}, \bibinfo {author} {\bibfnamefont {S.}~\bibnamefont {Matsumoto}}, \
  and\ \bibinfo {author} {\bibfnamefont {T.~T.}\ \bibnamefont {Yanagida}},\
  }\href {\doibase 10.1103/PhysRevD.85.095011} {\bibfield  {journal} {\bibinfo
  {journal} {Phys. Rev. D}\ }\textbf {\bibinfo {volume} {85}},\ \bibinfo
  {pages} {095011} (\bibinfo {year} {2012})},\ \Eprint
  {http://arxiv.org/abs/1202.2253} {arXiv:1202.2253 [hep-ph]} \BibitemShut
  {NoStop}%
\bibitem [{\citenamefont {Arvanitaki}\ \emph {et~al.}(2013)\citenamefont
  {Arvanitaki}, \citenamefont {Craig}, \citenamefont {Dimopoulos},\ and\
  \citenamefont {Villadoro}}]{Arvanitaki:2012ps}%
  \BibitemOpen
  \bibfield  {author} {\bibinfo {author} {\bibfnamefont {A.}~\bibnamefont
  {Arvanitaki}}, \bibinfo {author} {\bibfnamefont {N.}~\bibnamefont {Craig}},
  \bibinfo {author} {\bibfnamefont {S.}~\bibnamefont {Dimopoulos}}, \ and\
  \bibinfo {author} {\bibfnamefont {G.}~\bibnamefont {Villadoro}},\ }\href
  {\doibase 10.1007/JHEP02(2013)126} {\bibfield  {journal} {\bibinfo  {journal}
  {JHEP}\ }\textbf {\bibinfo {volume} {02}},\ \bibinfo {pages} {126} (\bibinfo
  {year} {2013})},\ \Eprint {http://arxiv.org/abs/1210.0555} {arXiv:1210.0555
  [hep-ph]} \BibitemShut {NoStop}%
\bibitem [{\citenamefont {Arkani-Hamed}\ \emph {et~al.}(2012)\citenamefont
  {Arkani-Hamed}, \citenamefont {Gupta}, \citenamefont {Kaplan}, \citenamefont
  {Weiner},\ and\ \citenamefont {Zorawski}}]{Arkani-Hamed:2012fhg}%
  \BibitemOpen
  \bibfield  {author} {\bibinfo {author} {\bibfnamefont {N.}~\bibnamefont
  {Arkani-Hamed}}, \bibinfo {author} {\bibfnamefont {A.}~\bibnamefont {Gupta}},
  \bibinfo {author} {\bibfnamefont {D.~E.}\ \bibnamefont {Kaplan}}, \bibinfo
  {author} {\bibfnamefont {N.}~\bibnamefont {Weiner}}, \ and\ \bibinfo {author}
  {\bibfnamefont {T.}~\bibnamefont {Zorawski}},\ }\href@noop {} {\  (\bibinfo
  {year} {2012})},\ \Eprint {http://arxiv.org/abs/1212.6971} {arXiv:1212.6971
  [hep-ph]} \BibitemShut {NoStop}%
\bibitem [{\citenamefont {Steinhardt}\ and\ \citenamefont
  {Turner}(1983)}]{Steinhardt:1983ia}%
  \BibitemOpen
  \bibfield  {author} {\bibinfo {author} {\bibfnamefont {P.~J.}\ \bibnamefont
  {Steinhardt}}\ and\ \bibinfo {author} {\bibfnamefont {M.~S.}\ \bibnamefont
  {Turner}},\ }\href {\doibase 10.1016/0370-2693(83)90727-X} {\bibfield
  {journal} {\bibinfo  {journal} {Phys. Lett. B}\ }\textbf {\bibinfo {volume}
  {129}},\ \bibinfo {pages} {51} (\bibinfo {year} {1983})}\BibitemShut
  {NoStop}%
\bibitem [{\citenamefont {Axenides}\ \emph {et~al.}(1983)\citenamefont
  {Axenides}, \citenamefont {Brandenberger},\ and\ \citenamefont
  {Turner}}]{Axenides:1983hj}%
  \BibitemOpen
  \bibfield  {author} {\bibinfo {author} {\bibfnamefont {M.}~\bibnamefont
  {Axenides}}, \bibinfo {author} {\bibfnamefont {R.~H.}\ \bibnamefont
  {Brandenberger}}, \ and\ \bibinfo {author} {\bibfnamefont {M.~S.}\
  \bibnamefont {Turner}},\ }\href {\doibase 10.1016/0370-2693(83)90586-5}
  {\bibfield  {journal} {\bibinfo  {journal} {Phys. Lett. B}\ }\textbf
  {\bibinfo {volume} {126}},\ \bibinfo {pages} {178} (\bibinfo {year}
  {1983})}\BibitemShut {NoStop}%
\bibitem [{\citenamefont {Linde}(1985)}]{Linde:1985yf}%
  \BibitemOpen
  \bibfield  {author} {\bibinfo {author} {\bibfnamefont {A.~D.}\ \bibnamefont
  {Linde}},\ }\href {\doibase 10.1016/0370-2693(85)90436-8} {\bibfield
  {journal} {\bibinfo  {journal} {Phys. Lett. B}\ }\textbf {\bibinfo {volume}
  {158}},\ \bibinfo {pages} {375} (\bibinfo {year} {1985})}\BibitemShut
  {NoStop}%
\bibitem [{\citenamefont {Seckel}\ and\ \citenamefont
  {Turner}(1985)}]{Seckel:1985tj}%
  \BibitemOpen
  \bibfield  {author} {\bibinfo {author} {\bibfnamefont {D.}~\bibnamefont
  {Seckel}}\ and\ \bibinfo {author} {\bibfnamefont {M.~S.}\ \bibnamefont
  {Turner}},\ }\href {\doibase 10.1103/PhysRevD.32.3178} {\bibfield  {journal}
  {\bibinfo  {journal} {Phys. Rev. D}\ }\textbf {\bibinfo {volume} {32}},\
  \bibinfo {pages} {3178} (\bibinfo {year} {1985})}\BibitemShut {NoStop}%
\bibitem [{\citenamefont {Kobayashi}\ \emph {et~al.}(2013)\citenamefont
  {Kobayashi}, \citenamefont {Kurematsu},\ and\ \citenamefont
  {Takahashi}}]{Kobayashi:2013nva}%
  \BibitemOpen
  \bibfield  {author} {\bibinfo {author} {\bibfnamefont {T.}~\bibnamefont
  {Kobayashi}}, \bibinfo {author} {\bibfnamefont {R.}~\bibnamefont
  {Kurematsu}}, \ and\ \bibinfo {author} {\bibfnamefont {F.}~\bibnamefont
  {Takahashi}},\ }\href {\doibase 10.1088/1475-7516/2013/09/032} {\bibfield
  {journal} {\bibinfo  {journal} {JCAP}\ }\textbf {\bibinfo {volume} {09}},\
  \bibinfo {pages} {032} (\bibinfo {year} {2013})},\ \Eprint
  {http://arxiv.org/abs/1304.0922} {arXiv:1304.0922 [hep-ph]} \BibitemShut
  {NoStop}%
\bibitem [{\citenamefont {Akrami}\ \emph {et~al.}(2020)\citenamefont {Akrami}
  \emph {et~al.}}]{Planck:2018jri}%
  \BibitemOpen
  \bibfield  {author} {\bibinfo {author} {\bibfnamefont {Y.}~\bibnamefont
  {Akrami}} \emph {et~al.} (\bibinfo {collaboration} {Planck}),\ }\href
  {\doibase 10.1051/0004-6361/201833887} {\bibfield  {journal} {\bibinfo
  {journal} {Astron. Astrophys.}\ }\textbf {\bibinfo {volume} {641}},\ \bibinfo
  {pages} {A10} (\bibinfo {year} {2020})},\ \Eprint
  {http://arxiv.org/abs/1807.06211} {arXiv:1807.06211 [astro-ph.CO]}
  \BibitemShut {NoStop}%
\bibitem [{\citenamefont {Linde}(1991)}]{Linde:1991km}%
  \BibitemOpen
  \bibfield  {author} {\bibinfo {author} {\bibfnamefont {A.~D.}\ \bibnamefont
  {Linde}},\ }\href {\doibase 10.1016/0370-2693(91)90130-I} {\bibfield
  {journal} {\bibinfo  {journal} {Phys. Lett. B}\ }\textbf {\bibinfo {volume}
  {259}},\ \bibinfo {pages} {38} (\bibinfo {year} {1991})}\BibitemShut
  {NoStop}%
\bibitem [{\citenamefont {Kofman}\ \emph {et~al.}(1994)\citenamefont {Kofman},
  \citenamefont {Linde},\ and\ \citenamefont {Starobinsky}}]{Kofman:1994rk}%
  \BibitemOpen
  \bibfield  {author} {\bibinfo {author} {\bibfnamefont {L.}~\bibnamefont
  {Kofman}}, \bibinfo {author} {\bibfnamefont {A.~D.}\ \bibnamefont {Linde}}, \
  and\ \bibinfo {author} {\bibfnamefont {A.~A.}\ \bibnamefont {Starobinsky}},\
  }\href {\doibase 10.1103/PhysRevLett.73.3195} {\bibfield  {journal} {\bibinfo
   {journal} {Phys. Rev. Lett.}\ }\textbf {\bibinfo {volume} {73}},\ \bibinfo
  {pages} {3195} (\bibinfo {year} {1994})},\ \Eprint
  {http://arxiv.org/abs/hep-th/9405187} {arXiv:hep-th/9405187} \BibitemShut
  {NoStop}%
\bibitem [{\citenamefont {Kasuya}\ \emph {et~al.}(1997)\citenamefont {Kasuya},
  \citenamefont {Kawasaki},\ and\ \citenamefont {Yanagida}}]{Kasuya:1996ns}%
  \BibitemOpen
  \bibfield  {author} {\bibinfo {author} {\bibfnamefont {S.}~\bibnamefont
  {Kasuya}}, \bibinfo {author} {\bibfnamefont {M.}~\bibnamefont {Kawasaki}}, \
  and\ \bibinfo {author} {\bibfnamefont {T.}~\bibnamefont {Yanagida}},\ }\href
  {\doibase 10.1016/S0370-2693(97)00809-5} {\bibfield  {journal} {\bibinfo
  {journal} {Phys. Lett. B}\ }\textbf {\bibinfo {volume} {409}},\ \bibinfo
  {pages} {94} (\bibinfo {year} {1997})},\ \Eprint
  {http://arxiv.org/abs/hep-ph/9608405} {arXiv:hep-ph/9608405} \BibitemShut
  {NoStop}%
\bibitem [{\citenamefont {Kawasaki}\ \emph {et~al.}(2013)\citenamefont
  {Kawasaki}, \citenamefont {Yanagida},\ and\ \citenamefont
  {Yoshino}}]{Kawasaki:2013iha}%
  \BibitemOpen
  \bibfield  {author} {\bibinfo {author} {\bibfnamefont {M.}~\bibnamefont
  {Kawasaki}}, \bibinfo {author} {\bibfnamefont {T.~T.}\ \bibnamefont
  {Yanagida}}, \ and\ \bibinfo {author} {\bibfnamefont {K.}~\bibnamefont
  {Yoshino}},\ }\href {\doibase 10.1088/1475-7516/2013/11/030} {\bibfield
  {journal} {\bibinfo  {journal} {JCAP}\ }\textbf {\bibinfo {volume} {11}},\
  \bibinfo {pages} {030} (\bibinfo {year} {2013})},\ \Eprint
  {http://arxiv.org/abs/1305.5338} {arXiv:1305.5338 [hep-ph]} \BibitemShut
  {NoStop}%
\bibitem [{\citenamefont {Kawasaki}\ and\ \citenamefont
  {Sonomoto}(2018)}]{Kawasaki:2017kkr}%
  \BibitemOpen
  \bibfield  {author} {\bibinfo {author} {\bibfnamefont {M.}~\bibnamefont
  {Kawasaki}}\ and\ \bibinfo {author} {\bibfnamefont {E.}~\bibnamefont
  {Sonomoto}},\ }\href {\doibase 10.1103/PhysRevD.97.083507} {\bibfield
  {journal} {\bibinfo  {journal} {Phys. Rev. D}\ }\textbf {\bibinfo {volume}
  {97}},\ \bibinfo {pages} {083507} (\bibinfo {year} {2018})},\ \Eprint
  {http://arxiv.org/abs/1710.07269} {arXiv:1710.07269 [hep-ph]} \BibitemShut
  {NoStop}%
\bibitem [{\citenamefont {Kawasaki}\ \emph {et~al.}(2018)\citenamefont
  {Kawasaki}, \citenamefont {Sonomoto},\ and\ \citenamefont
  {Yanagida}}]{Kawasaki:2018qwp}%
  \BibitemOpen
  \bibfield  {author} {\bibinfo {author} {\bibfnamefont {M.}~\bibnamefont
  {Kawasaki}}, \bibinfo {author} {\bibfnamefont {E.}~\bibnamefont {Sonomoto}},
  \ and\ \bibinfo {author} {\bibfnamefont {T.~T.}\ \bibnamefont {Yanagida}},\
  }\href {\doibase 10.1016/j.physletb.2018.05.014} {\bibfield  {journal}
  {\bibinfo  {journal} {Phys. Lett. B}\ }\textbf {\bibinfo {volume} {782}},\
  \bibinfo {pages} {181} (\bibinfo {year} {2018})},\ \Eprint
  {http://arxiv.org/abs/1801.07409} {arXiv:1801.07409 [hep-ph]} \BibitemShut
  {NoStop}%
\bibitem [{\citenamefont {Cheung}\ \emph {et~al.}(2012)\citenamefont {Cheung},
  \citenamefont {Elor},\ and\ \citenamefont {Hall}}]{Cheung:2011mg}%
  \BibitemOpen
  \bibfield  {author} {\bibinfo {author} {\bibfnamefont {C.}~\bibnamefont
  {Cheung}}, \bibinfo {author} {\bibfnamefont {G.}~\bibnamefont {Elor}}, \ and\
  \bibinfo {author} {\bibfnamefont {L.~J.}\ \bibnamefont {Hall}},\ }\href
  {\doibase 10.1103/PhysRevD.85.015008} {\bibfield  {journal} {\bibinfo
  {journal} {Phys. Rev. D}\ }\textbf {\bibinfo {volume} {85}},\ \bibinfo
  {pages} {015008} (\bibinfo {year} {2012})},\ \Eprint
  {http://arxiv.org/abs/1104.0692} {arXiv:1104.0692 [hep-ph]} \BibitemShut
  {NoStop}%
\end{thebibliography}%

\end{document}